\documentclass[12pt, twoside]{article}
\usepackage{amsmath,amsthm,amssymb,dsfont,bm}
\usepackage{times}
\usepackage{enumerate}

\def\titlerunning#1{\gdef\titrun{#1}}
\makeatletter
\def\author#1{\gdef\autrun{\def\and{\unskip, }#1}\gdef\@author{#1}}
\def\address#1{{\def\and{\\\hspace*{18pt}}\renewcommand{\thefootnote}{}%
\footnote {#1}}%
\markboth{\autrun}{\titrun}}
\makeatother
\def\email#1{e-mail: #1}
\def\subjclass#1{{\renewcommand{\thefootnote}{}%
\footnote{\emph{Mathematics Subject Classification (2010):} #1}}}
\def\keywords#1{\par\medskip
\noindent\textbf{Keywords.} #1}

\newtheorem{thm}{Theorem}[section]
\newtheorem{cor}[thm]{Corollary}
\newtheorem{lem}[thm]{Lemma}

\theoremstyle{definition}

\numberwithin{equation}{section}

\DeclareMathOperator{\tr}{tr}

\newcommand{\scal}[1]{\left\langle #1 \right\rangle}

\DeclareMathOperator{\dist}{dist}

\DeclareMathOperator{\ran}{ran}
\DeclareMathOperator{\spac}{spac}

\DeclareMathOperator{\distt}{dist}

\DeclareMathOperator{\rank}{rank}
\DeclareMathOperator{\supp}{supp}
\DeclareMathOperator{\disc}{disc}
\DeclareMathOperator{\res}{res}

\DeclareMathOperator{\loc}{loc}

\DeclareMathOperator{\proj}{pr}

\DeclareMathOperator{\spa}{sp}
\DeclareMathOperator{\W}{W}
\DeclareMathOperator{\M}{M}
\DeclareMathOperator{\Loc}{Loc}

\DeclareMathOperator{\spc}{sp}
\DeclareMathOperator{\ess}{ess}
\DeclareMathOperator*{\essinf}{ess\,inf}

\newcommand{\Z}{\mathbb{Z}}
\newcommand\what{\widehat}

\newcommand{\wt}{\widetilde}

\newcommand{\mce}{\mathcal E}

\newcommand{\veps}{\varepsilon}
\newcommand{\id}{\ensuremath{\mathds{1}}}

\newcommand{\mell}{\mathcal L}
\newcommand{\ieps}{I_\veps}
\renewcommand\P{\mathbb P}
\newcommand\E{\mathbb E}

\newcommand{\C}{{\mathbb C}}
\newcommand{\R}{{\mathbb R}}
\newcommand{\N}{{\mathbb N}}
\newcommand\beq{\begin{equation}}
\newcommand\eeq{\end{equation}}
\newcommand{\abs}[1]{\left\lvert #1 \right\rvert}
\newcommand{\norm}[1]{\left\lVert #1 \right\rVert}
\newcommand{\set}[1]{\left\{ #1 \right\}}
\newcommand{\pa}[1]{\left( #1 \right)}

\begin{document}

%%%%% To ease editing, add:

\baselineskip=17pt

%%%%%%%%%%%%%%%%

%% In the running head, give an abbreviation of the title. 
\titlerunning{Level spacing for continuum random Schr\"odinger operators}

\title{Level spacing and Poisson statistics for continuum random Schr\"odinger operators}

\author{Adrian Dietlein
\and 
Alexander Elgart}

\date{}

\maketitle

\address{
Adrian Dietlein: 
Mathematisches Institut,
  Ludwig-Maximilians-Universit\"at M\"unchen,
  Theresienstra\ss{e} 39,
  80333 M\"unchen, Germany; \email{dietlein@math.lmu.de}
\and
Alexander Elgart: 
Department of Mathematics,
Virginia Tech,
McBryde Hall,
225 Stanger Street,
Blacksburg, VA 24061, USA; \email{aelgart@vt.edu}}


%%%%%%%%

\begin{abstract}
We prove a probabilistic level-spacing estimate at the bottom of the spectrum for continuum alloy-type random Schr\"odinger operators, assuming  sign-definiteness of a single-site bump function and absolutely continuous randomness. More precisely, given a finite-volume restriction of the random operator onto a box of linear size $L$, we prove that with high probability the eigenvalues below some threshold energy $E_{\spc}$ keep a distance of at least $e^{-(\log L)^\beta}$ for sufficiently large $\beta>1$. This implies simplicity of the spectrum of the infinite-volume operator below $E_{\spc}$. Under the additional assumption of Lipschitz-continuity of the single-site probability density we also prove a Minami-type estimate and Poisson statistics for the point process given by the unfolded eigenvalues around a reference energy $E$.
%% Keywords are optional
\end{abstract}

\section{Introduction}
This work deals with spectral properties of random Schr\"odinger operators (RSO) $H_\omega=H_o+V_\omega$ acting on the Hilbert space $L^2(\R^d)$. Here $H_o$ is a fixed self-adjoint and non-random operator, for instance the Laplacian $-\Delta$, and $V_\omega$ is a real-valued multiplication operator whose spatial profile depends on a random variable $\omega$ from a probability space $(\Omega,\P)$. The interest in studying the properties of such operators was sparked by the seminal work of P. W. Anderson \cite{And}, who proposed the lattice counterpart of $H_\omega$ as a prototypical model for a metal-insulator transition. Specifically, he considered the operator $H^A_{\omega}:= -\Delta+V_\omega$ on $\ell^2(\Z^d)$, with random potential $V_\omega(x)=\lambda \omega_x$, $x\in\Z^d$. Here, the  $(\omega_x)_{x\in\Z^d}$ are a family of independent random variables distributed according to the uniform distribution on an interval.

For 'typical' configurations $\omega$ Anderson gave a semi-empirical argument supporting existence of a localized and a delocalized spectral regime for $H^A_{\omega}$ if $d\geq 3$. The localized spectral regime consists of pure point spectrum with exponentially localized eigenfunctions which cannot spread spatially under the dynamical evolution. Conversely, the delocalized spectral regime consists of wide-spread  eigenfunctions  which can carry diffusive transport. 

This model and its various extensions have since become focus of intensive research in both physics and mathematics. The effect of spectral localization due to disorder is relatively well understood by now  on a mathematical level, by virtue of  two known robust approaches to this phenomenon. In \cite{FS} Fr\"ochlich and Spencer developed a KAM-type method  known as the multiscale analysis, and in \cite{AM} Aizenman and Molchanov introduced the fractional moment method. We do  not attempt to give an exhaustive bibliography on the various extensions of those seminal works here but  refer to the recent monograph \cite{AW}. 

The folk wisdom in physics, and a frequently used litmus test for disordered systems, is that the spectral structure at energy $E$ is characterized by the limiting behavior of the point process of the appropriately rescaled eigenvalues around $E$. More precisely, for a large but finite box $\Lambda_L:=[-L/2,L/2]^d$ we consider the point process $\xi_{E,\omega}^L=\sum_{n} \delta_{L^d(E_{n,\omega}^L-E)}$, where $E_{n,\omega}^L$ are the eigenvalues of the finite-volume restriction of  the disordered system  $H_{\omega,L}$.

If the energy $E$ is within an exponentially localized spectral region, the eigenvalues localized in disjoint regions of space are almost independent. The point  process mentioned above is then expected to converge to a Poisson point process as the system's volume grows. Conversely, extended states imply that distant regions have mutual influence, leading to completely different eigenvalue statistics, such as the Gaussian orthogonal ensemble.  This duality is known as the spectral statistics conjecture. It plays an important role in the analysis of disordered systems, see e.g., \cite{Mirlin,ABF,EY}.

Poisson statistics were proved rigorously in the localization regime for the classical Anderson model $H_\omega^A$ in \cite{Min} and for a one-dimensional model in \cite{Molchanov}. The method from \cite{Min} is based on a probabilistic estimate on the event that two or more eigenvalues of $H_{\omega,L}$ are located in a small energy window. Such estimates are referred to as Minami estimates and have been further developed in \cite{BHS,GV,CGK,bourg,TV,HK}. However, with the exception of the one-dimensional case \cite{klopp}, these techniques heavily rely on the concrete structure of the random potential $V_\omega$ in $H^A_\omega$. In particular, they do not use the specific structure of kinetic energy and are only applicable for single-site potentials that are, or can be transformed to, rank-$1$ potentials (cf. the discussion in Section \ref{sec:OutlineProof} for more details). Our approach circumvents this difficulty by exploiting the kinetic energy term to find a sufficiently rich subset of the configuration space where the eigenvalues of $H_\omega$ are well spaced. We then invoke analytic estimates of Cartan type, developed earlier by Bourgain \cite{Bour} for an alternative approach towards Wegner's estimate, the key technical input of multiscale analysis. A similar analytic estimate was employed in the related paper \cite{IM}, where localization and level spacing for a specific lattice model with non-monotone rank-two random potential has been considered. This is however the only commonality of the two (\cite{IM} and ours) approaches.

One of our results is a Minami-type estimate for continuum random Schr\"odinger operators $H_\omega=-\Delta+V_\omega$ near the bottom ($=0$ without loss of generality) of the spectrum. Although this bound is much weaker than the usual Minami estimate known for $H_\omega^A$, it is sufficient to yield Poisson statistics for the point process of rescaled eigenvalues  of $H_\omega$. We now present an informal version  of this estimate (its precise statement will be formulated in Section \ref{sec1}). There exists $E_{\M}>0$ such that for all $K>0$ and sufficiently large $L\gg 1$ 
\beq
\label{IntroResult1}
\P\pa{\tr \id_{[E-\delta,E+\delta]}(H_{\omega,L}) \geq 2 } \leq C_K L^{4d}\delta |\log\delta|^{-K},
\eeq
provided that $\delta < 1$. This bound in turn is a consequence of our main technical result, a probabilistic estimate on the level spacing, i.e. the minimal distance between distinct eigenvalues (counting their multiplicities) of a self-adjoint operator in some spectral range. Informally, there exists $E_{\spc}>0$ such that 
\beq
\label{IntroResult2}
\P\Big(\sup_{E\leq E_{\spc}}  \tr\id_{[E-\delta,E+\delta]}(H_{\omega,L}) \geq 2 \Big)  \leq C L^{2d}  \exp\pa{-|\log\delta|^{1/(9d)}}
\eeq
for $L \gg 1$ and $\delta < 1$.
Beside the application to level statistics discussed above, the bound \eqref{IntroResult2} is also of independent interest. For instance, it allows to deduce simplicity of point spectrum below the energy $E_{\spc}$ (via the method in \cite{KM}). The level spacing is also expected to play an important role in the localization studies of an interacting electron gas in a random environment -- a subject of growing importance in theoretical and mathematical physics. In this context, the limited evidence from perturbative  \cite{AF1,AGKL,GMP1,AAB,Imb} approaches supports the persistence of a many-body localized phase for one-dimensional spin systems  in the presence of weak interactions. 

The paper is organized as follows: In Section \ref{sec1} we first introduce the model, a standard continuum random alloy-type Schr\"odinger operator, and discuss our  technical assumptions. We then present the main results and  outline their proofs. In Section \ref{sec2} we  formulate and prove some preparatory lemmas on clusters of eigenvalues. Sections \ref{sec:SpacingDeformedOp} and \ref{sec3}  contain the proofs of our two main results, Theorems \ref{thm:MainResult} and \ref{cor2}, that correspond to the informal  estimates \eqref{IntroResult1} -- \eqref{IntroResult2} above. These bounds yield statements on simplicity of spectrum and Poisson statistics for $H_\omega$ by known techniques \cite{CGK}; we outline the flow of these arguments in Section  \ref{sec4}.

\section{Model and results}
\label{sec1}

\subsection{Model} We consider a standard continuum alloy-type RSO
\beq
\label{def:TheOperator}
H_{\omega} := -\mu\Delta + V_\omega = - \mu\Delta + \sum_{k\in\Z^d} \omega_kV_k
\eeq
for $\mu>0$, acting on the Hilbert space $L^2(\R^d)$. Here $V_\omega$ is a random alloy-type potential with random coupling constants $\Omega\ni\omega=(\omega_k)_{k\in\Z^d}$ taken from a probability space $\pa{\Omega,\P}$ specified below. We now introduce technical assumptions on our model which we assume to hold for the rest of the section.
\vspace{0.3cm}
\begin{enumerate}
\item[$(V_1)$] The single-site bump functions $V_k$ are translates of a function $V_0$, $V_k(u)=V_0(u-k)$ for $u\in \R^d$ and $k\in\Z^d$. There exist constants $v_-, v_+ \in (0,1]$ and $r,R \in (0,\infty)$ such that 
\beq \label{eq:condV}
v_- \chi_{B_{r}(0)}\le V_0\le v_+\chi_{B_{R}(0)}. 
\eeq
\item[$(V_2)$] The random potential satisfies a covering condition: For constants $V_-,V_+\in(0,1]$ we have 
\beq
\label{eq:CovCond}
V_- \leq \sum_{k\in\Z^d} V_k \leq V_+.
\eeq
\item[$(V_3)$] The random couplings $\omega=(\omega_k)_{k\in\Z^d}\in \R^{\Z^d}$ distribution is given by  $\P:= \bigotimes_{\Z^d} P_0$. The single-site probability measure $P_0$ is absolutely continuous with respect to Lebesgue measure on $\R$. Its Lebesgue density $\rho\in L^\infty(\R)$ satisfies $\supp(\rho) \subseteq[0,1]$.
\end{enumerate}
\vspace{0.3cm}
The assumptions $v_+,V_+\leq 1$ and $\supp(\rho)\subset[0,1]$ are made for convenience. The covering condition from $(V_2)$ is necessary for Theorems \ref{cor2} and \ref{cor4} below, but not for the level spacing estimate, Theorem \ref{thm:MainResult}. One could also include more general background operators $H_o$ instead of $-\mu\Delta$. However, in contrast to the situation for the classical Anderson model $H_A$, the choice of $H_o$ is not arbitrary. For further comments we refer to the discussion in Section \ref{sec:OutlineProof}. On the other hand, the regularity assumption on $P_0$ in $(V_3)$ is the principal technical assumption here.

Before we state detailed versions of our results we introduce notation and review some well-known properties of the random operator introduced above. 
For a Borel-measurable set $A\subset\R^d$ let $\chi_A$ be the $L^2$-projection onto $A$. The finite-volume restriction of $H_\omega$ to an open set $U\subset\R^d$ is defined as
\beq
\label{def:FiniteVolume}
H_{\omega,U} := -\mu\Delta_U + \sum_{k\in\Z^d} \omega_k V^U_k;\quad V_k^U := \chi_U V_k,
\eeq
where $-\Delta_U$ is endowed with Dirichlet boundary conditions. Hence the random potential $V_{\omega}^U = \sum_{k\in\Z^d} \omega_k V_k^U$ may depend on random variables from a $R$-neighbourhood of $U$ and the random operators $H_{\omega,U_1}, H_{\omega,U_2}$ are independent if $\dist(U_1,U_2) > 2R$. Here, $\dist(A,B):= \inf \{|a-b|:\, a\in A, b\in B\}$ for $A,B\subset\R^d$ and $|x|:= \max_{i}|x_i|$ for $x\in\R^d$. This choice of the finite-volume random potential to some extend matters in the proof of Theorem \ref{cor2}.
By $\Lambda_L:=[-L/2,L/2]^d$ and $\Lambda_L(x):=x+\Lambda_L$ we denote the box of side-length $L$ centered at $0\in \R^d$ ($x\in\R^d$, respectively), and abbreviate $H_{\omega,L}:= H_{\omega,\Lambda_L}$. In the same vein we set $V_k^L:= V_k^{\Lambda_L}$ etc.

The first property we need is a bound on the probability of spectrum of $H_{\omega,L}$ in an interval $I$, known as Wegner's estimate. It was first proved for the classical Anderson model $H_A$ in \cite{Weg} and later generalized substantially due to its central role in multiscale analysis. For further references and more recent developments we refer to \cite{CGK1,RMV,K}. 
\begin{itemize}
\item[($\W$)]
For fixed $E>0$ there exists a constant $C_{\W} = C_{\W,E}$ such that
\beq
\label{def:Wegner}
\P\pa{ \tr \id_{I}(H_{\omega,L}) \geq 1 } \leq C_{\W} L^d |I|
\eeq
for intervals $I \subset [0,E]$.
\end{itemize}
This estimate in particular implies regularity of the integrated density of states. Due to ergodicity of $H_\omega$, almost surely (with respect to $\P$) the function 
\beq
\mathcal N (E) :=  \lim_{L\to\infty} L^{-d} \tr \id_{(-\infty,E]}(H_{\omega,L})
\eeq
is well-defined for all $E\in\R$ and is non-random \cite{CL,PF}. Wegner's estimate ensures that $\mathcal N$ is Lipschitz continuous and possesses a Lebesgue density $n:=\mathcal N'$, the density of states of $H_\omega$. 

The second property that we employ is exponential spectral localization, which for the model considered here is known to hold at the bottom of the spectrum.  Both methods to study this phenomenon that were mentioned in the introduction have been extended to continuum RSO, initially in \cite{CH,AENSS}. For recent developments and further references we refer to \cite{BK,EK,GHK}. We'll work with the technically slightly stronger output generated by fractional moment analysis. For $x\in\R^d$ let $\chi_x:= \chi_{x+\Lambda_1}$.
 
\begin{itemize}
\item[($\Loc$)] 
There exists $E_{\loc}>0$, $1>s>0$ and constants  $C_{\loc},m>0$ such that for all $E<E_{\loc}$ and all $x,y\in\R^d$
\begin{equation}
\label{def:LocFMB}
\sup_{U\subset\R^d}\E \left[ \norm{\chi_x R_{E}(H_{\omega,U}) \chi_y}^s \right] \leq C_{\loc} e^{-m |x-y|}
\end{equation}
for all $x,y\in\R^d$. Here the supremum in $U$ is over open and bounded sets and $R_z(A):= (A-z)^{-1}$ denotes the resolvent of an operator $A$ for $z\in \C\setminus\sigma(A)$.
\end{itemize}
In \cite{AENSS} the bound \eqref{def:LocFMB} is proved with a boundary-adapted distance function in the exponent. As noted there, for Hamiltonians without magnetic potentials \eqref{def:LocFMB} also holds true with the usual distance $| \cdot |$; see also \cite{MNSS}.

\subsection{Results}Let $E_{i,L}^\omega$, $i\in\N$, denote the eigenvalues of $H_{\omega,L}$ in ascending order. Here, and in the following, the eigenvalues are counted according to their multiplicity. To quantify the level spacing of the operator $H_{\omega,L}$ in an interval $I\subset\R$ we set
\beq
\label{def:SpacingFunction}
\spac_{I}(H_{\omega,L}) := \inf\set{|E_{i,L}^\omega - E_{j,L}^\omega|:\, i\neq j,\, E_{i,L}^\omega,E_{j,L}^\omega\in I }
\eeq 
and denote $\spac_{E}(H_{\omega,L}):=\spac_{(-\infty,E]}(H_{\omega,L})$ for any $E\in\R$. The function $\spac_{I}(H_{\omega,L})$ is, by Weyl's inequality \cite[Ch. 4, Thm. 3.17]{Kato}, continuous for an appropriate topology on $\Omega$ and therefore measurable. The first result of this paper is a probabilistic bound on the minimal spacing of eigenvalues below the energy
\beq
\label{def:EnergySpacing}
E_{\spc} :=  \frac{\mu \pi^2V_-}{2R^2(2R+1)^d v_+}.
\eeq
As far as dependence on $V_\omega$ is concerned, this threshold is certainly sub-optimal.  But, regardless of the choice of random potential, the method below is limited to $E_{\spc}\leq \lambda_{2}^{(N)}/2$, where $\lambda_{2}^{(N)}$ is the second eigenvalue of the Neumann Laplacian on $\supp(V_0)$ (provided that the boundary is sufficiently regular). This is related to the fact that the spectral projection of this operator onto $[0,\lambda_{2}^{(N)})$ is rank one which we use explicitly in our reduction scheme, Lemmas \ref{lem:GoodConf1}--\ref{lem:GoodConf2} below. However, one can still partially carry out this reduction for an arbitrary fixed interval $[0,E]$. In the discrete setting, this output is sufficient to establish a weaker result, namely compound Poisson statistics, \cite{HK}. We expect that an adaptation of the method to our context will show compound Poisson statistics for energies above $E_{\spc}$. 

We state two versions of the level spacing estimate. The first -- stronger -- estimate relies on localization but does not require any additional assumptions besides $(V_1)$--$(V_3)$ above.

\begin{thm}[Probabilistic level-spacing estimate, Version 1]
\label{thm:MainResult}
For a fixed energy $E < \min\set{E_{\spc},E_{\loc}}$ there exist $\mell_{\spc} =\mell_{\spc,E}, C_{\spc}=C_{\spc,E}$ such that 
\beq
\label{eq:MainResult}
\P\pa{\spac_{E}(H_{\omega,L}) < \delta} \leq C_{\spc} L^{2d}  \exp\pa{-|\log\delta|^{1/(9d)}}
\eeq
holds for $L\geq \mell_{\spc}$ and $\delta <  1$.
\end{thm}

An estimate such as \eqref{eq:MainResult} is typically used  (as in this paper) to derive spectral properties of systems that exhibit localization. However, it is reasonable to expect that the estimate itself should not rely on localization per se, as long as some disorder is present.  This is  the case for the classical Anderson model $H^A$, where the Minami estimate holds irrespective of localization. We corroborate this intuition in our second version of the level-spacing estimate. To this end, we will use the following additional assumption:

\vspace{0.3cm}
\begin{enumerate}
\item [$(V_4)$] The single-site probability density $\rho$ is Lipschitz-continuous and bounded below,
\beq
\mathcal K:=\sup_{\substack{x,y\in[0,1]\\x\neq y}}\frac{\abs{\rho(x)-\rho(y)}}{\abs{x-y}}<\infty \quad \text{ and } \quad \rho_-:=\min_{x\in[0,1]} \rho(x) >0.
\eeq
\end{enumerate}
\vspace{0.3cm}

\begin{thm}[Probabilistic level-spacing estimate, Version 2]
\label{thm:MainResultv2}
Assume that $(V_4)$ holds. For fixed $E\in(0,E_{\spc})$ and $K>0$ there exist $\mell_{\spc} =\mell_{\spc,E,K}, C_{\spc}=C_{\spc,E,K}$ such that 
\beq
\label{eq:MainResultv2}
\P\pa{\spac_{E}(H_{\omega,L}) < \delta} \leq C_{\spc} L^{2d}  |\log \delta|^{-K}
\eeq
holds for $L\geq \mell_{\spc}$ and $\delta <1$.
\end{thm}

In Section \ref{sec:SpacingNoLoc} the probabilistic level-spacing estimate \eqref{eq:MainResultv2} is in fact proved for the larger class of deformed random Schr\"odinger operators $H_\omega= H_o+V_\omega$, where $H_o = -\mu G\Delta G + V_o$. Here, $G,V_o$ are sufficiently nice periodic potentials where $V_o$ is small in norm and $G\geq G_->0$ for a constant $G_-$. This enlargement of the model, which does not alter the arguments but complicates notation, is necessitated by the proof of the Minami-type estimate, Theorem \ref{cor2} below. There we use deformed operators with $G=V^{-1/2}$ and $V_o=EV^{-1}$ as auxiliary operators. For a short description of this step we refer to Section \ref{sec:OutlineProof}.

Degenerate eigenvalues of Schr\"odinger operators are typically caused by symmetry. Randomness tends to break symmetry and accidental degeneracies in generic random models are expected to occur with probability zero. The first result on simplicity of RSO goes back to Simon \cite{Sim1}, who proved almost sure simplicity of the eigenvalues of the standard Anderson model $H^A$. In \cite{JL} the almost sure simplicity was extended to the singular spectrum of $H_A$. The simplicity of pure point spectrum was also derived for some other forms of random potential in the discrete case in \cite{NNS}.

Here, we use a different route to establish this assertion which goes back to Klein and Molchanov, \cite{KM}. Namely,  the level spacing estimate, together with the argument from \cite{KM,CGK}, yields simplicity of the pure-point spectrum of the infinite-volume operator $H_\omega$ below $\min\{E_{\spc},E_{\loc}\}$.

\begin{cor}[Eigenvalue simplicity]
\label{cor3}
The spectrum in  $[0,\min\{E_{\spc},E_{\loc}\}] \cap \sigma(H_\omega)$ almost surely only consists of simple eigenvalues.
\end{cor}

We continue with the Minami-type estimate, which we prove for energies below
\beq
\label{def:EMin}
E_{\M}:= \frac{E_{\spc}V_-}{V_+} = \frac{\mu \pi^2 V_-^2 }{2R^2(2R+1)^d V_+ v_+}.
\eeq
For its proof we employ Theorem \ref{thm:MainResultv2} although a similar result could be deduced by working with Theorem \ref{thm:MainResult}. This would result in a faster $\delta$-decay in \eqref{eq:Minest} below but possibly (depending on the size of $\mu$) restrict the energy range from $E_{\M}$ to $\min\set{E_{\M},E_{\loc}}$. We note that Assumption $(V_4)$ is required in the proof of Theorem \ref{cor2} below even if Theorem \ref{thm:MainResult} is used.

\begin{thm}[Minami-type estimate]
\label{cor2}
Assume that $(V_4)$ holds. For fixed $E_0<E_{\M}$ and $K>0$ there exist $\mell_{\M}=\mell_{\M,E_0,K},C_{\M}=C_{\M,E_0,K}>0$  such that the following holds. For $E\leq E_{0}$
\beq\label{eq:Minest}
\P\pa{\tr\id_{[E-\delta,E+\delta]}(H_{\omega,L}) \geq 2}\leq C_{\M} L^{4d} \delta |\log\delta|^{-K}
\eeq
holds for all $L\geq \mell_{\M}$ and $\delta < 1$.
\end{thm}

Theorem \ref{cor2} is sufficient to prove, with the method from \cite{Min,Molchanov,CGK}, that the point process given by the properly rescaled eigenvalues around some small energy $E$ weakly converges to a Poisson point process as $L\to\infty$. The point process of the rescaled eigenvalues of $H_{\omega,L}$ around a fixed reference energy $E\in\R$ is given by
\beq
\xi_{E,\omega}^{L}(B):= \tr\pa{\id_{E+L^{-d}B}(H_{\omega,L})}
\eeq
for bounded, Borel-measurable sets $B\subset\R$.

\begin{thm}[Poisson statistics]
\label{cor4}
Assume that $(V_4)$ holds. Let $E<\min\{E_{\M},E_{\loc}\}$ such that the integrated density of states $\mathcal{N}$ is differentiable at $E$, with derivative $\mathcal{N}'(E)= n(E)>0$. Then, as $L\to\infty$, the point process $\xi_{E,\omega}^{L}$ converges weakly to the Poisson point process on $\R$ with intensity measure $n(E)dx$.
\end{thm}

Under assumption $(V_4)$ it follows from \cite{DGHKM} that $n(E)>0$ for (Lebesgue-) almost every $E\in (0,\min\set{E_{\loc},V_-})$. Hence the conclusion of the theorem holds for almost every energy $E \in [0,E_{\M}]$.

\subsection{Outline of the proofs}
\label{sec:OutlineProof}

In this section we comment on the arguments pertaining to the proof of  Theorem \ref{thm:MainResult}. The principle ideas used to establish Theorem \ref{thm:MainResultv2} are similar to the ones discussed below. We also address the derivation of Theorem \ref{cor2}  from Theorem \ref{thm:MainResultv2}. We will not comment on the proofs of the applications, Corollary \ref{cor3} and Theorem \ref{cor4}, as they follow via the strategy developed earlier in \cite{KM,Min,CGK}.

The known strategies to obtain a Minami estimate rely on the fact that the random potential itself, i.e. the operator $V_\omega$, readily satisfies this bound. Combined with the rank-$1$ structure of single-site bump functions in $H_A$, this feature allows to prove a Minami estimate for an arbitrary choice of the non-random local operator $H_o$ in $H_A$. Therefore it does not come as a surprise that the method already breaks down for the dimer potential, where the single-site bump functions are translates of $u=\id_{\set{0,1}}$, a rank-$2$ operator. Consequently, the effect of the kinetic energy term $H_o$ has to be taken into account in order to prove a Minami-type estimate for, say, the dimer model.

Typically, degenerate eigenvalues are a manifestation of symmetry within the system. A 'typical' kinetic energy term on a generic domain, say the Laplace operator on a  box, only possesses  -- if any -- global symmetries. In contrast, independence at distance of the random potential ensures that the symmetries of the random potential -- if any -- are local. The idea now is to harness the random potential to destroy global symmetries of the kinetic energy and, in turn, to use the repulsion of the kinetic energy to destroy local symmetries. A qualitative implementation of this observation was employed in the works \cite{Sim1, NNS} and \cite{JL} to prove simplicity of point spectrum, respectively singular spectrum.

Utilizing Wegner's estimate and localization we first reduce the level-spacing estimate \eqref{eq:MainResult} to the analysis of small clusters of at most $\ell^d$ eigenvalues, $\ell \sim |\log \delta|^\gamma \ll L$, for some $\gamma<1$, which are separated from the rest of the spectrum by a small spectral gap of size $\delta\ll\veps\ll |\log\delta|^{-1}$.  
\\
For such a cluster we apply a Feynman-Hellman type estimate, Lemma \ref{lem:EigCluster}. The Feynman-Hellman theorem states that for self-adjoint operators $A,B$ and a one-parameter spectral family $s\to A+sB$ we have $\tr P_s B  =  \partial_s \bar E^s \tr P_s$, where $P_s$ denotes the projection onto a cluster of eigenvalues and $\bar E^s$ denotes the central energy, i.e. the arithmetic mean of the eigenvalues in the cluster. In Lemma \ref{lem:EigCluster} we show that a stronger statement holds under the assumption that the cluster is tightly concentrated around $\bar E^s$, namely that $P_s \pa{B - \partial_s \bar E^s} P_s$ is small in operator norm. \\
We next argue that low lying eigenvalues cannot cluster everywhere in the configuration space. Let's assume we have bad luck and the cluster of at most $\ell^d$ eigenvalues is tightly concentrated around its central energy for configurations in a small neighborhood of some $\omega_0\in\Omega$. We then apply Lemma \ref{lem:EigCluster} for every $k\in\Gamma_L$ to the spectral family $s\to H_{\omega_0,L}+sV_k$. As an output, we find that the tight concentration of the cluster originates from high amount of local symmetry. More precisely, for every $k\in\Gamma_L$ one of the following two scenarios applies: Either all eigenfunctions of the cluster have almost no mass on $\supp(V_k)$ or they form an almost orthogonal family when restricted to $\supp(V_k)$. Via a bracketing argument we utilize this to conclude that the central energy $\bar E^{\omega_0}$ of the cluster has to be $\gtrsim \lambda_2^{(N)}$, the second eigenvalue of the kinetic energy $H_o$ restricted to $\supp(V_k)$ with Neumann boundary.

After iterating this argument, we obtain that for a cluster of eigenvalues $\lesssim \lambda_2^{(N)}$ there exists a quite rich set of configurations for which the eigenvalues of the cluster are rather far apart from each other. Let $\omega_0$ be such a configuration. The spectral gap surrounding the cluster ensures that quantities such as the central energy and the local discriminant of the cluster, defined in \eqref{eq:GenEst4}, can be extended to complex analytic functions in a vicinity of $\omega_0$ which is roughly of linear size $\veps$. We can now use a version of Cartan's Lemma, Lemma \ref{cont:dimlem3}, to show that in a neighborhood of the good configuration the eigenvalues of the cluster are still spaced with high probability. After collecting all the probabilistic estimates performed along the lines of this argument one obtains Theorem \ref{thm:MainResult}.

For the proof of Theorem \ref{cor2}, let us for the moment assume that $\sum_{k\in\Z^d} V_k=1$.
The principle idea leading from Theorem \ref{thm:MainResult} to a local estimate is to clone the interval $J:=J_0:=[E-\delta,E+\delta]$ for which we want to prove \eqref{eq:Minest}. Let $\set{J_k}_{k=1}^K$ be $K$ disjoint intervals  of length $2\delta$ and such that $\dist(J_k,J_0)\lesssim K\delta \ll 1$. We now utilize that (in view of  $\sum_k V_k=1$) a shift $\set{\omega_k}_k\to\set{\omega_k+\veps}_k$ in the configuration space results in an energy shift by $\veps$. Together with the homogeneity of the single-site probability measures -- which is where the additional assumption $(V_4)$ enters -- it implies that
\beq\label{eq:prshift}
\P\pa{\spac_{J_0}(H_{\omega,L}) < \delta} \sim \P\pa{\spac_{J_k}(H_{\omega,L}) < \delta}.
\eeq
Summing both sides over $1\leq k\leq K$ then yields 
\beq\label{eq:prshift2}
\P\pa{\spac_{J_0}(H_{\omega,L}) < \delta} \lesssim  K^{-1} \P\pa{\spac_{E_{\spc}}(H_{\omega,L}) < \delta},
\eeq
by arguing that the events on the right hand side of \eqref{eq:prshift} are nearly disjoint.
By choosing $K=~(L^d\delta)^{-1}$ we ensure that $\dist(J_k,J_0)\lesssim L^{-d}$, which turns out to be a sufficient condition for \eqref{eq:prshift} to hold. On the other hand, this yields the additional factor of $\delta$ on the right hand side of \eqref{eq:prshift2} and allows us to apply Theorem \ref{thm:MainResultv2}  to finish the argument.

In order to remove the constraint $V=\sum_{k\in\Z^d} V_k=1$ we consider the auxiliary operator $\wt H^E_\omega:=V^{-1/2}\pa{H_\omega-E}V^{-1/2}$. This motivates the introduction of the larger class of deformed random Schr\"odinger operators in Section \ref{sec:SpacingDeformedOp} for which we prove Theorem \ref{thm:MainResultv2}, see Theorem \ref{thm:MainResultW}. We  then  repeat the line of arguments above to conclude that \eqref{eq:Minest} holds for the operator $\wt H_\omega^E$ at energy zero. Exploiting that the spectrum of $H_\omega$ around energy $E$ and the spectrum of $\wt H_\omega^E$ around energy zero are in good agreement, see Lemma \ref{lem:SpecCong} for details, we finally obtain the same estimate for $H_\omega$ around energy $E$.

\section{Clusters of eigenvalues}
\label{sec2}

For the whole section let $A$ be a self-adjoint operator on a separable Hilbert space $\mathcal{H}$ Moreover we denote by $I\subset\R$ the interval which contains the cluster of eigenvalues and by $\veps$ the size of a spectral gap around $I$, with
\beq
\label{eq:Const}
|I|\leq \frac{1}{2} \quad \text{ and } 0<\veps < \frac{1}{12}.
\eeq
Throughout the section we also assume that
\beq
\label{eq:GenEst1}
n:=\tr\pa{\id_I(A)} < \infty\quad \text{ and } \quad \distt \pa{I,\sigma(A)\setminus I} \geq 6 \veps
\eeq
holds. The explicit choice of numerical values in \eqref{eq:Const} and \eqref{eq:GenEst1} is not particularly important.

\subsection{A Feynman-Hellmann type estimate}
In this subsection we consider the one-parameter operator family 
\beq
\label{eq:GenEst2}
(-\veps,\veps)\ni s \mapsto A_s:= A+sB,
\eeq
where $B$ is a bounded and self-adjoint operator with $\|B\|\leq 1$. For the enlarged interval $I_\veps := I+(-\veps,\veps)$ the properties \eqref{eq:GenEst1} yield
\beq
\label{eq:GenEst3}
n=\tr\pa{\id_{\ieps}(A_s)} \quad \text{ and } \quad \distt \pa{I_\veps,\sigma(A_s)\setminus I_\veps} \geq 4 \veps
\eeq
for all $s\in (-\veps,\veps)$. For such $s$ let $E_{1}^s,...,E_{n}^s$ denote the eigenvalues of $A_s$ in $I_\veps$ counted with multiplicities.

For the arithmetic mean 
$\bar E^s := n^{-1} \sum_{i} E_{i}^s$
of the eigenvalues of $A_s$ in $\ieps$ the Hellmann-Feynman formula gives $\tr \id_{\ieps}(A_s)B  = n \partial_s \bar E_s$. The next lemma provides additional information under the assumption that the $n$ eigenvalues in $\ieps$ are moving as a small (in comparison to $\veps$) cluster in the coupling parameter $s$. For the rest of the section we use the notation $P_s:= \id_{I_\veps}(A_s)$ for $s\in (-\veps,\veps)$.

\begin{lem}\label{lem:EigCluster}
Let $0<\delta<\veps$. If we have that
\beq
\label{lem:EigClusterAspt}
 \sup_{s\in (-\veps,\veps)}\sup_{i=1,...,n} |E_{i}^s- \bar E^s| \leq \delta,
\eeq
then the following bound holds:
\beq
\label{lem:EigClusterConcl}
\sup_{s\in (-\veps,\veps)} \big\Vert P_s\big(B - \partial_s \bar E^s\big) P_s \big\Vert \leq  9\sqrt{\frac{\delta}{\veps}}.
\eeq
\end{lem}

In the proof of Lemma \ref{lem:EigCluster} we apply the following bounds which are, for convenience, proven at the end of this section. 

\begin{lem}\label{lem:TechEst}
For $s\in (-\veps,\veps)$ we have
\beq
\label{eq:PfLemEigCl1}
\|\partial_sP_s\| \leq \frac{1}{2\veps} \quad \text{ and } \quad \|\partial^2_sP_s\| \leq \frac{1}{\pi\veps^2}.
\eeq
If moreover \eqref{lem:EigClusterAspt} holds for $0<\delta<\veps$, then also 
\beq
\label{eq:PfLemEigCl4}
\big\Vert \partial_s^2 \big(P_s \big(A_s-\bar E^s\big)P_s\big)\big\Vert \leq \frac{7}{\veps}.
\eeq
\end{lem}

\begin{proof}[Proof of Lemma \ref{lem:EigCluster}]
Assumption \eqref{lem:EigClusterAspt} gives
\beq
\label{eq:PfLemEigCl2}
\| (A_s- \bar E^s)P_s \| \leq \delta.
\eeq
Let $T_s = P_s( A_s-\bar E^s) P_s$. Then differentiation of $T_s$, together with \eqref{eq:PfLemEigCl1} from Lemma \ref{lem:TechEst} and \eqref{eq:PfLemEigCl2}, yields
\begin{align}
\big\Vert P_s \big(B-\partial_s\bar E^s\big) P_s \big\Vert &\leq 2 \|\partial_s P_s\| \|( A_s-\bar E^s) P_s\| + \|\partial_s T_s\|\notag\\
&\leq \frac{\delta}{\veps}+\|\partial_s T_s\|. \label{eq:PfLemEigCl3}
\end{align}
The lemma follows if $\|\partial_s T_s\|=\max_{\phi\in \mathcal{H}}|\scal{\phi,(\partial_s T_s)(s_0)\phi}| \leq 8\sqrt{\delta/\veps}$ for all $s\in (-\veps,\veps)$. Assume by contradiction that there exists $s_0\in (-\veps,\veps)$ and a normalized $\psi \in \mathcal{H}$ such that
$|\scal{\psi,(\partial_s T_s)(s_0)\psi}| > 8\sqrt{\delta/\veps}$. Set $T_{s,\psi} := \scal{\psi, T_s\psi}$. Then  either 
$ (\partial_sT_{s,\psi})(s_0) > 8\sqrt{\delta/\veps}$ or $(\partial_s T_{s,\psi})(s_0) < -8\sqrt{\delta/\veps}$, and without loss of generality we can assume the former relation. 
Using the bound \eqref{eq:PfLemEigCl4} from Lemma \ref{lem:TechEst} we get that for $s_1\in (-\veps,\veps)$
\beq
(\partial_sT_{s,\psi})(s_1) \geq (\partial_sT_{s,\psi})(s_0) - \frac{7}{\veps} |s_1-s_0| \geq 8 \sqrt{\frac{\delta}{\veps}} -  \frac{7}{\veps} |s_1-s_0|
\eeq
by the fundamental theorem of calculus. 
Hence for any $s$ in 
\[\mathcal S:=\Big\lbrace s\in (-\veps,\veps):\ |s-s_0| \leq \frac{\sqrt{\delta\veps}}{2}\Big\rbrace\] 
we have $(\partial_sT_{s,\psi})(s) > 9\sqrt\delta/(2\sqrt\veps)$. It implies the existence of $s_2 \in \mathcal S$ such that
\beq
\delta \geq \abs{T_{s_2,\psi}} \geq  \frac{\sqrt{\delta\veps}}{2} \frac{9\sqrt\delta}{2\sqrt\veps}  -|T_{s_0,\psi}| \geq \frac{5}{4}\delta,
\eeq
a contradiction. 
\end{proof}

\begin{proof}[Proof of Lemma \ref{lem:TechEst}]  
Let $I_{+} = \sup I$ and $I_-=\inf I$. By $\gamma_{I,\veps}$ we denote the contour consisting of the oriented line segments $[I_- -3\veps +i\infty,I_- -3\veps -i\infty]$ and $[I_+ +3\veps -i\infty,I_+ +3\veps +i\infty]$. On $\ran(\gamma_{I,\veps})$ the resolvent of $A_s$ can be estimated as $\|R_{x+iy}(A_s)\| \leq ((2\veps)^2+y^2)^{-1/2}$ and hence
\begin{align}
\|\partial_s P_s\| &= \frac{1}{2\pi}\Big\Vert\int_{\gamma_{I,\veps}}dz\, R_z(A_s)B R_z(A_s) \Big\Vert\notag \\
&\quad\leq \frac{1}{\pi} \int_{\R}dy\, \frac{1}{(2\veps)^2+y^2} = \frac{1}{2\veps},\label{eq:lemTechBoundPf1}\\
\|\partial_s^2 P_s\| &= \frac{1}{\pi}\Big\Vert\int_{\gamma_{I,\veps}}dz\, R_z(A_s)B R_z(A_s)BR_z(A_s)\Big\Vert\notag\\
&\quad \leq \frac{2}{\pi} \int_{\R}dy\, \frac{1}{((2\veps)^2+y^2)^{3/2}} = \frac{1}{\pi\veps^2} \label{eq:lemTechBoundPf2}.
\end{align}
We next turn to estimate \eqref{eq:PfLemEigCl4}. For the rest of the proof we set $P:=P_s$, $\dot P := \partial_s P_s$ and $\ddot P:= \partial_s^2 P_s$ as well as $\bar E := \bar E^s$. We have 
\beq
\partial_s \big(P(A_s-\bar E)P\big) =\dot P P(A_s-\bar E)P+P(A_s-\bar E)P\dot P+P(B-\dot{\bar E})P.
\eeq
Taking the second derivative, we get
\begin{multline}\partial^2_s\pa{P(A_s-\bar E)P}  =\Big\{\Big(\ddot P(A_s-\bar E)P+\dot P^2(A_s-\bar E)P +\dot P P(B-\dot{\bar E})P\\ + \dot P P(A_s-\bar E)\dot P\Big)+h.c.\Big\} +\Big\{\dot P(B-\dot{\bar E})P+h.c.\Big\}-P\ddot{\bar E}P
\end{multline}
This yields
\begin{align}
\norm{\partial_s^2 \pa{P(A_s-\bar E)P}} &\leq 2 \|\ddot P \| \|(A_s-\bar E)P\| + 4\|\dot P \|^2 \|\pa{A_s-\bar E}P\| \notag\\
&\quad+ 8  \|\dot P\| +  |\ddot {\bar E}|,\label{eq:lemTechBoundPf3}
\end{align}
where we used $\|P\|=1$, $\|B\|\le 1$, and the fact that the first derivative of ${\bar E} = n^{-1} \tr\pa{PA_s}$ satisfies
\beq
-1\le \dot{\bar E} =\frac{1}{n}  \pa{  2\tr (P \dot  P A_s ) +\tr \pa{P B}} =  \frac{1}{n} \tr\pa{PB}\le 1.
\eeq
Using now the estimates \eqref{eq:PfLemEigCl1}, \eqref{eq:PfLemEigCl2}, and $\ddot{\bar E}= n^{-1} \tr\big(\dot P B\big)$, we obtain
\beq
\norm{\partial_s^2 \pa{P(A_s-\bar E)P}}\le 2 \frac{\delta}{\pi\veps^2} + 4 \frac{\delta}{4\veps^2} + \frac{4}{\veps}  + \frac{1}{2\veps}
\leq \frac{2\delta}{\veps^2} + \frac{5}{\veps}.
\eeq
\end{proof}

\subsection{The local discriminant and a Cartan estimate}

With the notation from the preceding section, if at least two eigenvalues of $A$ are inside $I$, $n \geq 2$, then we define the local discriminant of $A_s$ on $\ieps$ as
\beq
\label{eq:GenEst4}
\disc_{\ieps}(A_s) := \prod_{1\leq i<j\leq n} (E_{i}^s - E_{j}^s)^2
\eeq
for $s\in(-\veps,\veps)$.

\begin{lem}\label{lem:LocDisc}
The local discriminant, interpreted as a function $(-\veps,\veps) \ni s \mapsto \disc_{\ieps}(A_s)$, has an extension to a complex analytic function on $B_{3\veps}^\C:=\{z\in \C:\, |z| < 3\veps\}$ which is bounded by $1$.
\end{lem}

Let now $N\in\N$ and $0\leq B_k\leq 1$ be self-adjoint operators for $k=1,...,N$ such that $\sum_k B_k \leq 1$. We consider the $N$-parameter spectral family
\beq
(-\veps,\veps)^N \ni{\bf s}:= (s_1,..,s_N)\mapsto A+\sum_{k=1}^N s_k B_k.
\eeq
Then the following version of Cartan's lemma holds for the local discriminant.

\begin{lem}
\label{cont:dimlem3}
If for fixed $0<\delta_0<\veps$ there exists ${\bf s_0}\in (-\veps,\veps)^{N}$ such that 
\beq
\spac_{I_\veps}(A_{{\bf s_0}}) > \delta_0,
\eeq
then there exist constants $C_1,C_2$ (independent of all the relevant parameters above) such that
\beq
\label{Cont:dimlem3Stat}
\abs{\set{{\bf s} \in (-\veps,\veps)^N:\, \spac_{I_\veps}(A_{{\bf s}}) < \delta}} \leq  C_1 N (2\veps)^N \exp\pa{-\frac{C_2}{n^2} \abs{\frac{\log \delta }{\log \delta_0 }}}
\eeq
for all $\delta\in (0,1)$.
\end{lem}

\begin{proof}[Proof of Lemma \ref{lem:LocDisc}] 
Thanks to \eqref{eq:GenEst1}, we have $\id_{\ieps}(A_s)= \id_{I_{3\veps}+i\R}(A_s)$ and $\id_{\ieps^c}(A_s)= \id_{I_{3\veps}^c+i\R}(A_s)$ for $s\in (-\veps,\veps)$. I.e. the two projections can be extended to the complex analytic operators
\begin{align}
B_{3\veps}^{\C} \ni s &\mapsto \id_{I_{3\veps}+i\R}(A_s),\\
B_{3\veps}^{\C} \ni s &\mapsto \id_{I_{3\veps}^c+i\R}(A_s),
\end{align}
defined via the holomorphic functional calculus, \cite{Kato}. Define
\beq
z\mapsto p_s(z) = \det \pa{\id_{I_{3\veps}}(A_s)(A_s-z)+\id_{I_{3\veps}^c}(A_s)} = \prod_{i=1}^n (E_{i}^s-z),
\eeq
which is a polynomial of degree $n$ in $z$. Here the $E_{i,s}$, $i=1,...,n$, are the eigenvalues of $A_s$ for $s\in(-3\veps,3\veps)$ counted with multiplicities. For fixed $z\in\C$ the function $s\mapsto p_s(z)$ can be extended to a complex analytic function $\widetilde p_s(z)$ on $B_{3\veps}^{\C}$, given by
\begin{align}
B_{3\veps}^{\C}\ni s \mapsto\widetilde p_s(z) &= \det \pa{\id_{I_{3\veps}+i\R}(A_s)(A_s-z)+\id_{I_{3\veps}^c+i\R}(A_s)}.
\end{align}
If we write the polynomial as $\widetilde p_{s}(z) = \sum_{k=0}^n a_{k}(s) z^k$, then the coefficients $a_k(s)$ are also complex analytic on $B_{3\veps}^{\C}$ since they can be expressed via evaluations of $\widetilde p_s(z)$ at different values of $z$, for instance via Lagrange polynomials. For $s\in B_{3\veps}^{\C}$ the resultant of $\widetilde p_s$ and $\widetilde p_s'$, which is a polynomial of degree $n(n-1)$ in each of the coefficients $a_n(s)$, is then 
\beq
\label{eq:ContDisc}
\res(p_s,p_s')=(-1)^{n(n-1)/2}\prod_{i<j} \,(\lambda_i(s) - \lambda_j(s))^2,
\eeq
where the $\lambda_i(s)$ are an arbitrary enumeration of the zero's of $\widetilde p_s$. For $s\in(-\veps,\veps)$ this agrees, up to the prefactor $\pm 1$ in \eqref{eq:ContDisc} with the local discriminant $\disc_{I_{\veps}}(A_s)$ for $A_s$ defined above.
This proves the first part of the lemma. For the second part we note that the $\lambda_i(s)$ in \eqref{eq:ContDisc} are the eigenvalues of $A_s$ in $B_{3\veps}^\C$. Because $\sigma(A_s) \subset \sigma(A)+ B_{3\veps}^{\C}$ for $s\in B_{3\veps}^{\C}$, and because $|I|\leq 1/2$ and $\veps<1/12$, this shows that $|\lambda_i(s)-\lambda_j(s)| \leq 1$ holds for $s\in B_{3\veps}^\C$.
\end{proof}

\begin{proof}[Proof of Lemma \ref{cont:dimlem3}]
We define the map
\beq
\label{ContdimLem3Eq1}
(-\veps,\veps)^N \ni{\bf z}:= (z_1,..,z_N)\mapsto F(z) := \disc_{I_\veps}\Bigg( A+\sum_{k=1}^N z_k B_k\Bigg).
\eeq
Lemma \ref{lem:LocDisc} implies that for $\xi=\pa{\xi_i}_{i}\in [-1,1]^{N}$ the map
\beq
\label{ContdimLem3Eq2}
(-\veps,\veps)\ni s \mapsto F(s\xi_1,...,s\xi_N)
\eeq
can be extended to a complex analytic map on $B_{3\veps}^{\C}$. If we set  
$F_\veps (z) := F(2\veps z)$ for $z\in [-1/2,1/2]^N$ then $[-1/2,1/2]\ni s \mapsto F_\veps(s\xi_1,...,s\xi_N)$ is real analytic and can be extended to a complex analytic map on $B_{3/2}^\C$ with $|F_\veps|\leq 1$. Since by assumption there exists $z_0\in [-1/2,1/2]^N$ such that $|F_\veps(z_0)| > \delta_0^{n^2}$ Lemma 1 from \cite{Bour} is applicable and yields
\beq
\label{ContdimLem3Eq3}
\abs{\set{z\in[-1/2,1/2]^N:\, |F_\veps(z)| < \delta}} \leq C_1 N \exp\pa{- \frac{C_2}{ n^2}\abs{\frac{\log \delta}{\log \delta_0}}}  
\eeq
for $\delta\in(0,1)$ and constants $C_1,C_2$ that are uniform in all relevant parameters. Estimate \eqref{Cont:dimlem3Stat} now follows from \eqref{ContdimLem3Eq3} and
\begin{align}
\abs{\set{{\bf s}\in (-\veps,\veps)^N:\, \spac_{I_\veps}(A_{\bf s}) < \delta}}
&\leq\abs{\set{{\bf s}\in (-\veps,\veps)^N:\, \disc_{I_\veps}(A_{\bf s}) < \delta}}\label{ContdimLem3Eq4}\\
&= \abs{\set{{\bf s}\in(-\veps,\veps)^N:\, |F({\bf s})| < \delta}}\notag\\
&= (2\veps)^N \abs{\set{z\in[-1/2,1/2]^N:\, |F_\veps(z)| < \delta}} \notag.
\end{align}
\end{proof}

\section{Proof of the level spacing estimates}
\label{sec:SpacingDeformedOp}

In this section we prove Theorems \ref{thm:MainResult} and \ref{thm:MainResultv2}. In the proof of Theorem \ref{cor2} we have to apply Theorem \ref{thm:MainResultv2} for the auxiliary operators $\wt H_\omega^E$ described in Section \ref{sec:OutlineProof}. In order to prove Theorem \ref{thm:MainResultv2} and simultaneously establish the same result for the auxiliary operators, we prove a variant of Theorem \ref{thm:MainResult} for the deformed random Schr\"odinger operators $-\mu G \Delta G + V_o +V_\omega,$ where $G,V_o$ are bounded $\Z^d$-periodic potentials.

In the course of this section we denote both, the standard RSO and the deformed RSO, by $H_\omega$. To absorb this ambiguity of notation we specify the setup for each subsection separately.

\subsection{Existence of good configurations}
\label{sec:GoodConf}

In this section we work with the deformed random Schr\"odinger operators
\beq
\label{eq:wtH}
H_\omega :=  -\mu G \Delta G + V_o +V_\omega.
\eeq
Here $G,V_o$ are bounded and $\Z^d$-periodic potentials and $V_\omega=\sum_{k\in\Z^d} \omega_kV_k$ is as introduced in Section \ref{sec1}. In particular, the properties $(V_1)$--$(V_3)$ still hold. Moreover, we assume that $G$ satisfies $ G_-\leq G \leq G_+$ with constants $G_-,G_+ \in (0,\infty)$. 

The first step towards Theorems \ref{thm:MainResult} and \ref{thm:MainResultv2} is to prove that the configuration space $\Omega$ contains a sufficiently rich set of configurations for which the energy levels are well-spaced. More precisely, let $\omega_0\in\Omega$ and assume that a cluster of eigenvalues is isolated from the rest of the spectrum by a gap. Then the lemma below shows that there exists at least one configuration close to $\omega_0$ such that the cluster literally separates into clusters consisting of single eigenvalues. The lemma states that if localization for the cluster of eigenvalues is known then the amount of random variables that is needed to obtain such a 'good configuration' can be reduced to $\ell^d \ll L^d$. If localization is not known then the lemma can still be applied for $\ell=L$, see Lemma \ref{lem:GoodConfv2} below. 

We first introduce some additional notation. For $L>0$ let $\Gamma_L:= \Lambda_{L+R} \cap \Z^d$ be the index set of relevant couplings for the operator $H_{\omega,L}$ and for $x \in \Lambda_L$ let $\Gamma_{\ell,x}:= \Gamma_L\cap \Lambda_\ell(x)$, where the dependence on $L$ is suppressed in notation. In the same vein we denote by $\omega_{0,\Lambda_\ell(x)}$ and $\omega_{0,\Lambda^c_\ell(x)}$ the restrictions of $\omega_0\in[0,1]^{\Gamma_L}$ to the index sets $\Gamma_{\ell,x}$, respectively $\Gamma_L\setminus \Gamma_{\ell,x}$. We also define the local subcubes $Q^{\Lambda_\ell(x)}_{\veps}(\omega_0):= \omega_{0,\Lambda_\ell(x)}+[-\veps,\veps]^{\Gamma_{\ell,x}}$ for $\veps>0$. Moreover, for $\omega_1 \in [0,1]^{\Gamma_L}$ we set
\begin{align}
\label{def:ModBox}
Q^{\pa{x,\ell}}_\veps(\omega_1,\omega_0)&:=\omega_{1,\Lambda^c_\ell(x)}\times Q^{\Lambda_\ell(x)}_{\veps}(\omega_0)\notag\\
&:=\set{\omega=(\omega_{1,\Lambda^c_\ell(x)},\omega_{\Lambda_\ell(x)})\in[0,1]^{\Gamma_L}:\, \omega_{\Lambda_\ell(x)}\in Q^{\Lambda_\ell(x)}_{\veps}(\omega_0)}.
\end{align}
For $n\in\N$, $L\geq \ell>0$ and $r>0$ we define
\beq
\xi_{L,\ell,n,r} :=  \frac{\mu \pi^2 G_-^2}{2R^2(2R+1)^d v_+}\pa{V_- -v_+L^de^{-m \ell} -26\sqrt{n}\ell^{-r}}-\|V_o\|.
\eeq
\begin{lem}
\label{lem:GoodConf1}
Let $0<\veps<1/12$, $r>0$ and $m>0$ be fixed. Moreover, let $L\geq \ell \geq (8n)^{1/(2d+2r)}$ and $\omega_0,\omega_1\in [0,1]^{\Gamma_L}$ such that the following holds:
\begin{enumerate}
\item[(i)] $\omega_{1,\Lambda_\ell(x)} \in Q_{\veps}^{\Lambda_\ell(x)}(\omega_0)$.
\item[(ii)]
There exist eigenvalues $E^{\omega_1}_1\leq...\leq E^{\omega_1}_n \leq \xi_{L,\ell,n,r}$ of $H_{\omega_1,L}$ which are separated from the rest of the spectrum: For the cluster $\mathcal{C}_n^{\omega_1}:= \{E^{\omega_1}_1,...,E^{\omega_1}_n\}$ we have
\beq
\label{eq:LocEstAs1}
\dist\pa{\mathcal C_n^{\omega_1},\sigma(H_{\omega_1,L})\setminus \mathcal C_n^{\omega_1}}\ge 8\veps.
\eeq
\item[(iii)]
the spectral projection $P_{\omega_1}$ of $ H_{\omega_1,L}$ onto the cluster $\mathcal{C}_n^{\omega_1}$ is localized with localization center $x\in \Lambda_L$, i.e.
\beq\label{eq:locP}
\norm{P_{\omega_1} \id_{\Lambda_1(y)}}\le e^{-m \ell}
\eeq 
for all $y\in\Lambda_L$ that satisfy $\abs{x-y}>\ell$.
\end{enumerate}
Then there exists $\widehat\omega \in Q^{\pa{x,\ell}}_{\veps}(\omega_1,\omega_0)$ such that
\beq
\label{eq:LocEstConc}
\min_{i=1,...,n-1}|E^{\what\omega}_{i+1}-E^{\what\omega}_i| > 8\veps \ell^{-(n-1)(2d+2r)}.
\eeq
Here, $E_1^\omega\leq ... \leq E_n^\omega$ for $\omega\in Q^{\pa{x,\ell}}_{\veps}(\omega_1,\omega_0)$ denote the ascendingly ordered eigenvalues of $H_{\omega,L}$ in the interval $[E_1^{\omega_1}-2\veps,E_n^{\omega_1}+2\veps]$.
\end{lem}

Up to an iterative step, this lemma is a consequence of the following assertion.

\begin{lem}
\label{lem:GoodConf2}
Assume that the assumptions of Lemma \ref{lem:GoodConf1} hold. Then there exists $\widehat\omega \in Q^{\pa{x,\ell}}_{\veps-\veps \ell^{-(2d+2)}}(\omega_1,\omega_0)$ and $1\leq k \leq n-1$ such that 
\beq
\label{eq:dimLocEstConc}
E^{\what \omega}_{k+1}-E^{\what \omega}_{k} > 8\veps \ell^{-(2d+2r)}.
\eeq
\end{lem}

\begin{proof}[Proof of Lemma \ref{lem:GoodConf2}]
We set $I:=[E_1^{\omega_1}-\veps,E_n^{\omega_1}+\veps]$, where the dependence of $I$ on $\veps$ is suppressed in notation. By Weyl's inequality on the movement of eigenvalues and assumption \eqref{eq:LocEstAs1} we can without  loss of generality assume that 
\beq
\dist(I,\sigma(H_{\omega_0,L})\setminus I) \geq 6\veps.
\eeq
If this was not true, then \eqref{eq:dimLocEstConc} would readily hold. Another application of Weyl's inequality yields $\tr \id_{I_{\veps}}(H_{\omega,L}) = n$ for $\omega\in Q^{\pa{x,\ell}}_{\veps}(\omega_1,\omega_0)$, where $I_{\veps}:= I+[-\veps,\veps] = [E_1^{\omega_1}-2\veps,E_n^{\omega_1}+2\veps]$. This justifies the notation 
$E^\omega_1\leq...\leq E^\omega_n$ for the ascendingly ordered eigenvalues of $H_{\omega,L}$ in the interval $I_\veps$. For such $\omega$ we also define $\bar E^{\omega} := n^{-1}\sum_{i=1}^{n} E^{\omega}_i$. For notational convenience we set $Q:= Q^{\pa{x,\ell}}_{\veps}(\omega_1,\omega_0)$. We now assume that
\beq 
\label{eq:AsptContra'}
\max_{\omega\in Q}\max_{i=1,\ldots,n} |E^\omega_i-\bar E^\omega| \leq 8 n \veps \ell^{-(2d+2)}
\eeq 
holds. For fixed $k\in \Gamma_{\ell,x}$ there exists $-\veps<a_k<\veps$ such that $\omega_1 + e_k \pa{a_k + (-\veps,\veps)} \subset Q$. Here $e_k$ is the unit vector onto $k\in\Gamma_{\ell,x}$. Hence Lemma \ref{lem:EigCluster} can be applied to the operator family 
\beq
(-\veps,\veps) \ni s \mapsto H_{\omega_1+e_k a_k,L}+s V^L_k
\eeq
for $\delta= 8n\veps \ell^{-(2d+2)}$.
For $P_\omega:=\id_{I_{\veps}}(H_{\omega,L})$ let 
\beq\label{eq:alpha}
\alpha^{\omega_1}_k:=  (\partial_{\omega_k}\bar E^\omega)(\omega_1)=\frac{1}{n}\tr P_{\omega_1}V^L_k \geq 0,
\eeq 
where we have used the Hellmann-Feynman theorem. Evaluation of \eqref{lem:EigClusterConcl} at $s=-a_k$ yields the bound 
\beq 
\label{eq:LocEq2'}
\norm{P_{\omega_1}\pa{V^L_k-\alpha^{\omega_1}_k}P_{\omega_1}} \leq  26 \sqrt{n} \ell^{-d-r}
\eeq
for every $k\in\Gamma_{\ell,x}$. We next decompose $\Gamma_{\ell,x}$ into disjoint subsets $(U_t)_{t\in\mathcal T}$ such that $|k-l|>2R$ holds for $k,l\in U_t$, $k\neq l$, and such that $|\mathcal T|\leq (2R+1)^d$. For the sets $\Lambda_R^L(k):= \Lambda_R(k) \cap \Lambda_L$, $k\in \Gamma_L$, we then have $\Lambda_R^L(k) \cap \Lambda_R^L(k') = \emptyset$ for $k,k' \in U_t$ with $k\neq k'$. For fixed $t\in\mathcal T$
Neumann decoupling hence yields
\beq
\label{eq:LocEq0}
\tr P_{\omega_1} H_{\omega_1,L} \geq \sum_{k\in U_t} \tr P_{\omega_1}G\big(-\mu\Delta^{(N)}_{\Lambda_R^L(k)} \big)G - n\|V_o\|,
\eeq
where we also used that $V_k \omega_{1,k} \geq 0$ for all $k\in U_t\subset\Gamma_{\ell,x}$. After summing \eqref{eq:LocEq0} over $t\in\mathcal T$, we obtain
\beq
\label{eq:LocEq4'}
\tr P_{\omega_1} H_{\omega_1,L} \geq (2R+1)^{-d} \sum_{k\in\Gamma_{\ell,x}} \tr P_{\omega_1}G\big(-\mu\Delta^{(N)}_{\Lambda_R^L(k)} \big)G - n\|V_o\|.
\eeq 
Since $\Lambda_R^L(k)$ is a hyperrectangle with side-lengths bounded by $R$, we have
\beq
-\Delta_{\Lambda^L_R(k)}^{(N)} \geq \frac{\pi^2}{R^2}R_k ,
\eeq
where $R_k$ is the projection onto $\ran(\Delta^{(N)}_{\Lambda^L_R(k)})$. With the shorthand notation \[C_{\omega_1,k}:= G\chi_{\Lambda_R^L(k)}P_{\omega_1}\chi_{\Lambda_R^L(k)}G\] we conclude that 
\beq
\eqref{eq:LocEq4'} \geq \frac{\mu\pi^2}{R^2(2R+1)^d}  \sum_{k\in\Gamma_{\ell,x}} \tr C_{\omega_1,k}R_k- n\|V_o\|.
\eeq 
Next, we bound the trace on the right hand side as
\beq
\tr C_{\omega_1,k}R_k=\tr C_{\omega_1,k}-\tr C_{\omega_1,k}(\chi_{\Lambda_R^L(k)}-R_k)\ge \tr C_{\omega_1,k}-\norm{C_{\omega_1,k}}={\sum}' \nu_j,
\eeq
where $\pa{\nu_j}_j$ are the eigenvalues of $C_{\omega_1,k}$ counted with multiplicity and ${\sum}'$ stands for the sum of all but the largest eigenvalue of $C_{\omega_1,k}$. Here we also used that $\rank(\chi_{\Lambda_R^L(k)}-R_k)=1$. Since $\sigma(C_{\omega_1,k})\setminus \set{0}=\sigma(P_{\omega_1}\chi_{\Lambda_R^L(k)}G^2\chi_{\Lambda_R^L(k)}P_{\omega_1})\setminus \set{0}$ and, by \eqref{eq:LocEq2'}, 
\beq
P_{\omega_1}\chi_{\Lambda_R^L(k)}G^2\chi_{\Lambda_R^L(k)}P_{\omega_1}\ge \frac{G_-^2}{v_+} P_{\omega_1}V^L_kP_{\omega_1}\ge \frac{G_-^2}{v_+}\pa{\alpha^{\omega_1}_k -26 \sqrt{n} \ell^{-d-r}}P_{\omega_1},
\eeq 
we deduce by the min-max principle
\begin{align}
\tr C_{\omega_1,k}R_k &\ge {\sum}' \nu_j\ge  \frac{1}{v_+}\pa{\alpha^{\omega_1}_k - 26 \sqrt{n} \ell^{-d-r}}  \pa{\tr P_{\omega_1}-1}\notag\\
&=\frac{(n-1)G_-^2}{v_+}\pa{\alpha^{\omega_1}_k - 26 \sqrt{n} \ell^{-d-r}}.
\end{align}
This implies that
\beq
\tr P_{\omega_1} H_{\omega_1,L} \geq \frac{\mu \pi^2 G_-^2(n-1)}{R^2(2R+1)^d v_+}\sum_{k\in\Gamma_{\ell,x}} \pa{\alpha_k^{\omega_1} -26 \sqrt{n} \ell^{-d-r}} - n\|V_o\|.
\eeq 
Moreover, \eqref{eq:locP} and \eqref{eq:alpha} yield
\beq
\label{eq:LocEq3'}
\sum_{k\in\Gamma_{\ell,x}}\alpha^{\omega_1}_k=\frac{1}{n}\sum_{k\in\Gamma_{\ell,x}} \tr P_{\omega_1}V^L_k \geq \frac{1}{n}\sum_{k\in\Gamma_L} \tr P_{\omega_1}V^L_k- v_+  L^d e^{-m\ell}.
\eeq
Now we can use $\sum_{k\in\Gamma_L} \tr P_{\omega_1}V_k^L \geq n V_-$. Putting all bounds together, we get
\begin{align}
\bar E^{\omega_1} = \frac{1}{n} \tr P_{\omega_1} H_{\omega_1,L} &\geq \frac{\mu \pi^2 G_-^2}{2R^2(2R+1)^d v_+}\pa{V_- -v_+L^de^{-m \ell} -26\sqrt{n}\ell^{-r}}-\|V_o\|\notag\\
&= \xi_{L,\ell,n,r}.
\end{align}
\end{proof}

\begin{proof}[Proof of Lemma \ref{lem:GoodConf1}]
First, we  apply Lemma \ref{lem:GoodConf2} to the cluster $\mathcal{C}_n^{\omega_0}=\{E^{\omega_0}_1,...,E^{\omega_0}_n\}$ and the set $Q_0:=Q^{(x,\ell)}_{\veps}(\omega_1,\omega_0)$ in configuration space. We conclude that there exists $\omega_{0,2} \in Q_1:=Q^{(x,\ell)}_{\veps-\veps \ell^{-(2d+2r)}}(\omega_1,\omega_0)$ and $1\leq k_1 \leq n-1$ such that 
\beq
E^{\omega_{0,2}}_{k_1+1}-E^{\omega_{0,2}}_{k_1} > 8 \veps \ell^{-(2d+2r)}.
\eeq
If $k_1=1$ or $k_1=n-1$ then we isolated one eigenvalue from the rest of the eigenvalues and only proceed with one cluster of eigenvalues. In the other cases we obtain two sets of eigenvalues $E^{\omega_1}_1\leq...\leq E_{k_1}^{\omega_{0,2}}$ and $E^{\omega_{0,2}}_{k_1+1}\leq...\leq E^{\omega_{0,2}}_n$ which both satisfy \eqref{eq:LocEstAs1} for $\veps_1:=\veps \ell^{-(2d+2r)}$. We then apply Lemma \ref{lem:GoodConf2} to the set of eigenvalues $E^{\omega_{0,2}}_1\leq...\leq E_{k_1}^{\omega_{0,2}}$. This yields $\omega_{0,3} \in Q_2:= Q^{(x,\ell)}_{\veps_1-\veps_1 \ell^{-(2d+2r)}}(\omega_1,\omega_{0,2})$ and $1\leq k_2\leq k_1-1$ such that
\beq 
E^{\omega_{0,3}}_{k_2+1}-E^{\omega_{0,3}}_{k_2} > 8\veps_1 \ell^{-(2d+2r)}.
\eeq
Set $\veps_2:=\veps_1 \ell^{-(2d+2r)}$. Then, since $|\omega_2-\omega_1|_\infty \leq \veps_1-\veps_2$ we have
\beq
E^{\omega_{0,3}}_{k_1+1}-E^{\omega_{0,3}}_{k_1} > 8\veps_1- 2 \pa{\veps_1-\veps_2} \geq 8 \veps_2
\eeq
by Weyl's inequality and we can apply Lemma \ref{lem:GoodConf2} to the set $E^{\omega_{0,3}}_{k_1+1}\leq...\leq E^{\omega_{0,3}}_n$ of eigenvalues. Overall we found $\omega_{0,4}\in Q_3:= Q_{\veps_2-\veps_2 \ell^{-(2d+2r)}}(\omega_1,\omega_{0,3})$ and up to four clusters of eigenvalues which are separated from each other (and the rest of the spectrum of $H_L$) by $8 \veps_3 := 8 \veps_2 \ell^{-(2d+2r)}$. We repeat this procedure at most $n-1$ times until each cluster consists of exactly one eigenvalue.
\end{proof}

\subsection{Proof of Theorem \ref{thm:MainResultv2}}
\label{sec:SpacingNoLoc}

The setup is as in Section \ref{sec:GoodConf}, i.e. 
\beq
H_\omega :=  -\mu G \Delta G + V_o +V_\omega
\eeq
and $G,V_o,V_\omega$ satisfy the conditions specified there. Let
\beq
E_{\spc}:= \frac{\mu \pi^2 V_- G_-^2}{2R^2(2R+1)^d v_+} - \|V_o\|.
\eeq
Next is this section's main result, which for $G=\id_{L^2(\R^d)}$ gives Theorem \ref{thm:MainResultv2}.

\begin{thm}\label{thm:MainResultW}
Assume that $(V_4)$ holds. Then for fixed $E\in(0,E_{\spc})$ and $K>0$ there exist constants $\mell_{\spc} =\mell_{\spc,E,K}, C_{\spc}=C_{\spc,E,K}$ such that 
\beq
\label{eq:MainResultv2W}
\P\pa{\spac_{E}(H_{\omega,L}) < \delta} \leq C_{\spc} L^{2d}  |\log \delta|^{-K}
\eeq
holds for $L\geq \mell_{\spc}$ and $\delta < 1$.
\end{thm}

 In order to extract \eqref{eq:Minest} at energy $E$ from \eqref{eq:MainResultv2W} we have to apply the estimate multiple times for the $E$-dependent  potential $V_o=EV^{-1}$ and for a set of slightly varying $L$-dependent coupling constants $\mu_L$. This is why we will occasionally comment  in the sequel on the stability of constants as functions of  $V_o$ and $\mu$ variables. 

Besides the existence of good configurations for clusters of eigenvalues established above, the second ingredient for the proof of Theorem \ref{thm:MainResultv2} is a probabilistic estimate on the maximal size of generic clusters of eigenvalues. For lattice models, such estimates follow from an adaption of the method developed in \cite{CGK}, see \cite{HK}. The following assertion extends this idea.

\begin{lem}
\label{lem:Apriori}
For fixed $E>0$ and $\theta,\vartheta\in(0,1)$ there exist constants $c_\theta = c_{\theta,E},C_\vartheta=C_{\vartheta,E}>0$ such that
\beq
\P\big(\tr\id_{I}(H_{\omega,L})> c_{\theta }|I|^{-\theta}\big) \leq C_\vartheta L^{2d} |I|^{2-\vartheta}
\eeq
holds for all intervals $I\subset (-\infty,E]$.
\end{lem}

%\begin{remark}
%A slight adaption of the proof shows the following. For fixed $\vartheta\in(0,1)$, $\gamma\in[1,\infty)$ and $E>0$ there exists a constant $C_{\vartheta,\gamma}=C_{\vartheta,\gamma,E}$ such that for all $M\geq 1$ 
%\beq
%\P\pa{\tr\id_{I}(H_{\omega,L})> M} \leq C_{\vartheta,\gamma} L^{2d} |I|^{1-\vartheta} \max\set{|I|,  M^{-\gamma}}
%\eeq
%holds for all intervals $I\subset (-\infty,E]$.
%\end{remark}

\begin{proof}
As in the proof of Lemma \ref{lem:AprioriBound}, we apply Lemma \ref{lem:SpecCong} to estimate for a fixed interval $I := E_0+[-\delta G_-^{-1} ,\delta G_-^{-1}]\subset (-\infty,E]$
\beq
\label{eq:someEq1}
\tr \id_{I}(H_{\omega,L}) \leq \tr \id_{[-\delta ,\delta ]}(\wt H_{\omega,L}),
\eeq
where $\wt H_{\omega}:= -\mu\Delta+  G^{-2}(V_o-E_0) + G^{-2}V_\omega$. Then \eqref{eq:someEq1} implies 
\beq
\label{eq:someEq2}
\P\big(\tr\id_{I}( H_{\omega,L})> C \big) \leq \P\big(\tr\id_{[-\delta,\delta]}(\wt H_{\omega,L})> C \big)
\eeq
for any $C>0$. By $\xi(\mce,\wt H_{\omega,L}^{\omega_x=0},\wt H_{\omega,L}^{\omega_x=1})\geq 0$ we denote the the spectral shift function at energy $\mce$ of the operators
\beq
\label{eq:Aprioribound1}
\wt H_{\omega,L}^{\omega_x=0} := \wt H_{\omega,L} - \omega_x G^{-2} V_x \quad \text{ and } \quad \wt H_{\omega,L}^{\omega_x=1} := \wt H_{\omega,L} + (1-\omega_x) G^{-2} V_x.
\eeq
We then define the random variable
\beq
\label{eq:Aprioribound2}
X_\omega:= \sup_{x\in\Gamma_L} \essinf_{\mce\in [-\delta,\delta]} \xi(\mce,\wt H_{\omega,L}^{\omega_x=0},\wt H_{\omega,L}^{\omega_x=1}) \geq 0,
\eeq
where $\Gamma_L:= \Lambda_{L+R}\cap \Z^d$. Because $X_\omega$ is integer valued, we have
\begin{align}
\label{eq:Aprioribound3}
&\P\big(\tr\id_{[-\delta,\delta]}(\wt H_{\omega,L})> X_\omega\big)\notag \\
&\qquad\leq \E \big[\tr\id_{[-\delta,\delta]}(\wt H_{\omega,L}) (\tr\id_{[-\delta,\delta]}(\wt H_{\omega,L})- X_\omega) \id_{\set{\tr\id_{[-\delta,\delta]}(\wt H_{\omega,L})> X_\omega}} \big].
\end{align}
Omitting the $\omega,L$-subscripts for the moment, we get for $\mce\in[-\delta,\delta]$ and $x\in\Gamma_L$
\begin{align}
\tr \id_{[-\delta,\delta]}(\wt H) &= \tr\big(\id_{(-\infty, \delta]}(\wt H) - \id_{(-\infty,\mce]}( \wt H)\big) + \tr\big(\id_{(-\infty,\mce]}(\wt H) - \id_{(-\infty,-\delta]}(\wt H)\big)\notag\\
&\leq \tr\big(\id_{(-\infty, \delta]}(\wt H^{\omega_x=0}) -\id_{(-\infty, \mce]}(\wt H^{\omega_x=0})\big)\notag \\
&\quad + \tr\big(\id_{(-\infty,\mce]}(\wt H^{\omega_x=0}) - \id_{(-\infty, \mce]}(\wt H)\big)\notag\\
&\quad +\tr\big(\id_{(-\infty,\mce]}(\wt H^{\omega_x=1}) - \id_{(-\infty,-\delta]}(\wt H^{\omega_x=1})\big)\notag\\
&\quad +\tr\big(\id_{(-\infty,\mce]}(\wt H)-\id_{(-\infty,\mce]}(\wt H^{\omega_x=1})\big)\notag\\
&\leq \tr \id_{[-\delta,\delta]}(\wt H^{\omega_x=0}) + \tr \id_{[-\delta,\delta]}(\wt H^{\omega_x=1})\notag\\
&\quad + \xi(\mce,\wt H^{\omega_x=0},\wt H^{\omega_x=1}).\label{eq:Aprioribound4}
\end{align}
Since the inequality holds for all $\mce \in [-\delta,\delta]$ we obtain
\beq
\label{eq:Aprioribound5}
\tr \id_{[-\delta,\delta]}(\wt H) \leq  \tr \id_{[-\delta,\delta]}(\wt H^{\omega_x=0}) + \tr \id_{[-\delta,\delta]}(\wt H^{\omega_x=1}) + X.
\eeq
Next we use \eqref{eq:Aprioribound5} to estimate \eqref{eq:Aprioribound3}. We first note that for a constant $C'_{\W}$ the Wegner estimate
\beq
\label{WegnerWithoutX}
\E\big[ \tr \id_{[-\delta,\delta]}(\wt H_{\omega,L}^{\omega_x=1}) \big] \leq C'_{\W} L^d \delta,
\eeq
holds, for instance via \cite{CGK1} or \cite{K}. With \eqref{WegnerWithoutX} at hand we obtain
\begin{align}
\label{eq:Aprioribound6}
\eqref{eq:Aprioribound3} &\leq V_-G_+^{2} \sum_{x\in\Gamma_L} \E\left[ \tr G^{-2}V_x \id_{[-\delta,\delta]}(\wt H_{\omega,L}) \tr \id_{[-\delta,\delta]}(\wt H_{\omega,L}^{\omega_x=0})  \right] \notag\\
&\quad+ V_- G_+^{2} \sum_{x\in\Gamma_L}  \E\left[ \tr G^{-2} V_x  \id_{[-\delta,\delta]}(\wt H_{\omega,L})\tr \id_{[-\delta,\delta]}(\wt H_{\omega,L}^{\omega_x=1}) \right] \notag\\
&\leq C_{\vartheta} (2\delta)^{2-\vartheta} L^{2d}.
\end{align}
In the last inequality we applied the Birman-Solomyak formula \cite{BS} to obtain
\beq
\int_{[0,1]}d\omega_x\, \tr G^{-2} V_x  \id_{[-\delta,\delta]}(\wt H_{\omega,L}) = \int_{[-\delta,\delta]}d\mce\, \xi(\mce, \wt H_{\omega,L}^{\omega_x=0},\wt H_{\omega,L}^{\omega_x=1}).
\eeq
The estimate then follows from the local $L^p$-boundedness of the spectral shift function as a function in energy \cite{CHN}, applied for $p=\vartheta^{-1}$.

We finish the argument by proving the upper bound $X_{\omega} \leq c_{\theta}|I|^{-\theta}$, where $c_\theta$ does not depend on $\omega$. After estimating $X_\omega$ as
\begin{align}
\label{eq:Aprioribound9}
X_\omega &\leq \sup_{x\in\Gamma_L} \frac{1}{2\delta} \int_{[-\delta,\delta]} d\mce\, \xi(\mce, \wt H_{\omega,L}^{\omega_x=0},\wt H_{\omega,L}^{\omega_x=1})\notag\\
&\leq \sup_{x\in\Gamma_L} (2\delta)^{-\theta} \Big(\int_{[-\delta,\delta]} d\mce\, \xi(\mce, \wt H_{\omega,L}^{\omega_x=0},\wt H_{\omega,L}^{\omega_x=1})^{1/\theta}\Big)^\theta
\end{align}
we can again apply the local $L^p$-boundednes of the spectral shift function, this time for $p=1/\theta$, to obtain $X_\omega \leq c_\theta |I|^{-\theta}$. 
\end{proof}

Before we start proving Theorem \ref{thm:MainResultv2} we state a version of the 'good configurations Lemma' \ref{lem:GoodConf1} which is adapted to the present situation, i.e. $L=\ell$ and $r=d/2+1$. Let
\beq
\xi_{L,n} :=  \frac{\mu \pi^2 G^{2}_-}{2R^2(2R+1)^d v_+}\pa{V_- -26\sqrt{n}L^{-d-1}},
\eeq
where we have omitted the term $v_+L^de^{-mL}$, which does not appear in \eqref{eq:LocEq3'} in the $\ell=L$ case. The choice $r=d/2+1$ ensures that for $E_{\spc}-\xi_{L,n} \sim \sqrt{n}L^{-d/2-1} \leq  C_1 L^{-1}$, with $C_1$ as in Lemma \ref{lem:AprioriBound}.

\begin{lem}[Lemma \ref{lem:GoodConf1} for $\ell=L$, $r=d/2+1$]
\label{lem:GoodConfv2}
Let $0<\veps<1/12$, $L \geq 1$ and $\omega_0,\omega_1\in [0,1]^{\Gamma_L}$ such that the following holds:
\begin{enumerate}
\item[(i)] $\omega_{1} \in Q_{\veps}(\omega_0)$.
\item[(ii)]
There exist eigenvalues $E^{\omega_1}_1\leq...\leq E^{\omega_1}_n \leq \xi_{L,n}$ of $H_{\omega_1,L}$ which are separated from the rest of the spectrum: For the cluster $\mathcal{C}_n^{\omega_1}:= \{E^{\omega_1}_1,...,E^{\omega_1}_n\}$ we have
\beq
\label{eq:LocEstAs1'}
\dist\pa{\mathcal C_n^{\omega_1},\sigma(H_{\omega_1,L})\setminus \mathcal C_n^{\omega_1}}\ge 8\veps.
\eeq
\end{enumerate}
Then there exists $\widehat\omega \in Q_{\veps}(\omega_0)$ such that
\beq
\label{eq:LocEstConc'}
\min_{i=1,...,n-1}|E^{\what\omega}_{i+1}-E^{\what\omega}_i| > 8\veps L^{-(n-1)(3d+2)}.
\eeq
Here, $E_1^\omega\leq ... \leq E_n^\omega$ for $\omega\in Q_{\veps}(\omega_0)$ denote the ascendingly ordered eigenvalues of $H_{\omega,L}$ in the interval $[E_1^{\omega_1}-2\veps,E_n^{\omega_1}+2\veps]$.
\end{lem}

\begin{proof}[Proof of Theorem \ref{thm:MainResultW}]
For fixed $E\in (0,E_{\spc})$ we first decompose the interval $[-\|V_o\|,E]$ into a family $(K_i)_{i\in\mathcal I}$ of intervals with side length $|K_i| = \kappa< E_{\spc}$, with $|K_{i+1}\cap K_i| \geq \kappa/2$, and such that $|\mathcal I| \leq 2(E_{\spc}+\|V_o\|) \kappa^{-1} + 1$. Let $i\in\mathcal I$ and define $K_{i,8\veps} := K_i + [-8\veps,8\veps]$ for $\veps\in(0,1/12)$. Let $\theta\in (0,1)$. Then the probability of the event
\beq
\label{eq:DimerEq1}
\Omega_{i,\veps} := \big\{\tr \id_{K_i}(H_{\omega,L})\leq c_\theta |K_i|^{-\theta} \text{ and } \tr \id_{K_{i,8\veps}\setminus K_i}(H_{\omega,L}) = 0 \big\}
\eeq
can be estimated by Wegner's estimate and Lemma \ref{lem:Apriori} with $\vartheta=1/2$ as 
\beq
\P\pa{\Omega_{i,\veps}} \geq 1-16 C_{\W}L^d\veps - C L^{2d} \kappa^{3/2}.
\eeq 
For $0<\delta<\kappa/2$ this yields
\begin{align}
\label{eq:DimerEq2}
&\P\pa{\spac_{E}(H_{\omega,L})<\delta} \notag\\
&\qquad\leq \sum_{i\in\mathcal{I}} \P\pa{\set{\spac_{K_i}(H_{\omega,L})<\delta}\cap \Omega_{i,\veps}} + 16C_{\W} |\mathcal I| L^d \veps + C |\mathcal I | L^{2d} \kappa^{3/2}.
\end{align}
We next partition the configuration space $[0,1]^{\Gamma_L}$ into (not necessarily disjoint) cubes $Q_j$, $j\in\mathcal{J}$, of side length $2\veps$, i.e. $|Q_j| = (2\veps)^{|\Gamma_L|}$, such that 
\beq
|\mathcal{J}| \leq ((2\veps)^{-1}+1)^{|\Gamma_L|} \quad \text{ and } \quad \sum_{j\in\mathcal J}\P(Q_j) \leq 1+ 4 \veps |\Gamma_L|  \rho_+
\eeq 
hold. Now, fix $i\in \mathcal I$ and $j\in\mathcal J$ such that $Q_j\cap \Omega_{i,\veps}\neq\emptyset$, and let $\omega_{i,j} \in Q_j\cap \Omega_{i,\veps}$. This configuration satisfies
\begin{equation}
n_{i,j}:=\tr \id_{K_i}(H_{\omega_{i,j},L}) \leq c_\theta \kappa^{-\theta} \quad \text{ and } \quad \dist\pa{K_i,\sigma(H_{\omega_{i,j},L})\setminus K_i} \geq 8\veps.
\end{equation}
Due to the choice $r=d/2+1$ in Lemma \ref{lem:GoodConfv2}, we have $E<\xi_{L,L^d}$. Hence the lemma is applicable for sufficiently large $L$ and yields $\what\omega_{i,j} \in Q_j$ such that
\begin{equation}
\spac_{K_{i,\veps}}(H_{\what\omega_{i,j},L}) \geq 8\veps L^{-(n_{i,j}-1)(3d+2)}.
\end{equation}
This in turn can be used as an input for Lemma \ref{cont:dimlem3} with $\delta_0:= 8\veps L^{-(n_{i,j}-1)(3d+2)}$. For $Q_j =: \times_{k\in\Gamma_L} [a_{j,k},b_{j,k}]$ we obtain
\begin{align}
&\P\pa{Q_j\cap \big\{\spac_{K_{i,2\veps}}(H_{\omega,L}) < \delta\big\}}\notag \\
& \qquad\leq \Big(\prod_{k\in\Gamma_L}\sup_{x\in [a_{j,k},b_{j,k}]} \rho(x) \Big)  \big|\big\{ \omega\in Q_j:\, \spac_{K_{i,2\veps}}(H_{\omega,L}) < \delta\big\} \big|\notag\\
 &\qquad\leq C_1 \pa{1+\frac{\mathcal K 2\veps}{\rho_-}}^{|\Gamma_L|} L^d \P\pa{Q_j} \exp\pa{\frac{-c'_{\theta}\kappa^{2\theta}|\log \delta|}{|\log 8\veps|+ c''_\theta\kappa^{-\theta} \log L}}.\label{eq:DimerEq4}
\end{align}
Here we used that $n_{i,j} \leq c_\theta \kappa^{-\theta}$ and that $\rho$ satisfies $(V_4)$, which for $k\in\Gamma_L$ gives
\beq
\sup_{x\in [a_{j,k},b_{j,k}]} \rho(x) \leq \inf_{x\in [a_{j,k},b_{j,k}]} \rho(x) + \mathcal K 2\veps \leq \inf_{x\in [a_{j,k},b_{j,k}]} \rho(x) \pa{1+\frac{\mathcal K 2\veps}{\rho_-}}.
\eeq
The above estimate \eqref{eq:DimerEq4} holds for all pairs $i\in\mathcal I, j\in \mathcal J$ such that $Q_j\cap \Omega_{i,\veps}\neq \emptyset$. So far we assumed that $0<\veps<1/12$ and $0<\delta<\kappa/2<E_{\spc}/2$. If we set 
$\mathcal{J}_i := \{j\in\mathcal J:\, Q_j\cap \Omega_{i,\veps} \neq \emptyset\}$
for $i\in\mathcal I$, then
\begin{align}
\eqref{eq:DimerEq2} &\leq \sum_{i\in\mathcal I}\sum_{j\in\mathcal J_i}\P\pa{\set{\spac_{K_i}(H_{\omega,L})<\delta}\cap Q_{j}} + 16_{\W} |\mathcal I| L^d \veps + C |\mathcal I | L^{2d} \kappa^{3/2} \notag\\
&\leq C'_{\W} L^d \kappa^{-1} \veps + C' L^{2d} \kappa^{1/2} \notag\\
&\quad + C'_1 L^d \pa{1+\frac{\mathcal K 2\veps}{\rho_-}}^{|\Gamma_L|} (1+4\veps|\Gamma_L|\rho_+) \kappa^{-1}  \exp\pa{\frac{-c'_{\theta}\kappa^{2\theta}|\log \delta|}{|\log 8\veps|+ c''_\theta\kappa^{-\theta} \log L}}.  
\end{align}
For $0<\delta \leq \exp\pa{-(\log L)^5}$ we now choose 
\beq
\kappa:=|\log \delta|^{-1/(4\theta)} \quad \text{and} \quad \veps := \exp\pa{-|\log\delta|^{1/4}}.
\eeq 
Those choices in particular imply $\delta<\kappa/2$ for sufficiently large $L$.
Because $\veps |\Gamma_L| \leq 1$ for sufficiently large $L$ we end up with
\begin{align}
\P\pa{\spac_{E}(H_{\omega,L})< \delta} &\leq C_\theta'' L^{2d} |\log \delta|^{-1/(8\theta)} + C_1'' L^d  |\log \delta|^{1/(4\theta)} \exp\pa{-\tilde c_\theta |\log \delta|^{1/20}}\notag\\
&\leq C_{\spc} L^{2d} |\log \delta|^{-1/(8\theta)}
\end{align}
for a suitable constant $C_{\spc}$ and for $L\geq \mell_{\spc}$, where $\mell_{\spc}$ is sufficiently large.
\end{proof}

\subsection{Proof of Theorem \ref{thm:MainResult}}

For this section $H_\omega := -\mu\Delta+V_\omega$ denotes the standard random Schr\"odinger operator specified in Section \ref{sec1}. 

For the proof of Theorem \ref{thm:MainResult} we apply Lemma \ref{lem:GoodConf1} with two length scales $\ell\ll L$. The smaller scale $\ell$ serves two purposes. Together with localization it establishes a bound on the maximal size of clusters of eigenvalues that is stronger than the corresponding bound from Lemma \ref{lem:Apriori}. This is the reason why \eqref{eq:MainResult} is stronger than \eqref{eq:MainResultv2}. Secondly, we use the smaller scale $\ell$ to suppress the impact of the absolutely continuous density. This way we avoid the additional regularity assumption $(V_4)$ from Theorem \ref{thm:MainResultv2}.

For the scale $\mell_{\loc},m'$ as in Lemma \ref{cont:lem5} and $L\geq\ell\geq \mell_{\loc}$ we denote by $\Omega^{\loc}$ the set of $\omega\in\Omega$ that satisfy the following properties:\\
For all eigenpairs $(\lambda,\psi)$ of $H_{\omega,L}$ with $\lambda\in (-\infty,E_{\loc}]$ there exists $x\in\Lambda_L$ such that
\begin{enumerate}
\item[(i)] $\|\psi\|_y \leq e^{-m'\ell}$ for all $y\in\Lambda_{L}$ with $|x-y| \geq \ell+2R$,
\item[(ii)]$\dist\big(\sigma(H_{\Lambda^L_{2\ell+4R}(x)}),\lambda\big)\leq e^{-m'\ell}$,
\end{enumerate}
where we again use the notation $\Lambda_\ell^L(x) := \Lambda_\ell(x) \cap \Lambda_L$. According to the same lemma we have $\P\pa{\Omega^{\loc}} \geq 1-L^{2d}e^{-m'\ell}$. Moreover, we define for $\kappa>0$ 
\begin{align}\label{eq:pfThmInd1}
\Omega_{\kappa}^{\W} &:= \bigcap_{\substack{x,y\in\Lambda_L:\\ |x-y|>2\ell+6R}}\set{\distt\pa{\begin{array}{c}
\sigma(H_{\omega,\Lambda^L_{2\ell+4R}(x)})\cap (-\infty,E_{\loc}]\\ \text{ and } \\ \sigma(H_{\omega,\Lambda^L_{2\ell+4R}(y)})\cap (-\infty,E_{\loc}]
\end{array}} > 3 \kappa}, \notag\\
\Omega_{\kappa}^{g} &:= \Omega_{\kappa}^{\W}\cap\Omega^{\loc}.
\end{align}

If the Wegner estimate \eqref{def:Wegner} is applied to 'boxes' $\Lambda^L_{2\ell+4R}(x_1)$ and $\Lambda^L_{2\ell+4R}(x_2)$ with $\dist\pa{\Lambda^L_{2\ell+4R}(x_1),\Lambda^L_{2\ell+4R}(x_2)}>2R$, then the independence of the corresponding operators $H_{\omega,\Lambda^L_{2\ell+4R}(x_1)}$ and $H_{\omega,\Lambda^L_{2\ell+4R}(x_2)}$ yields
\beq
\P\pa{ \tr \id_{I}(H_{\omega,\Lambda^L_{2\ell+4R}(x_1)}) \geq 1 \text{ and } \tr \id_{I}(H_{\omega,\Lambda^L_{2\ell+4R}(x_2)})\geq 1} \leq C_{W}^{\prime 2} \ell^{2d} |I|^2
\eeq
for a slightly enlarged constant $C'_{\W}$. Together with Lemma \ref{cont:lem5} the probability of the event $\Omega_\kappa^g$ can be bounded from below by
\beq
\label{eq:ProbEstGoodSet}
\P\big(\Omega_{\kappa}^{g}\big) \geq 1-6 C^{\prime 2}_{\W} L^{2d} \ell^{2d} \kappa - L^{2d} e^{-m'\ell}
\eeq
for $L \geq \mell_{\loc}$, with $\mell_{\loc}$ as in Lemma \ref{cont:lem5}.

\begin{lem}
\label{cont:lem1}
Let $\mell_{\loc},m'$ as in Lemma \ref{cont:lem5}. Then, for $L\geq \ell \geq \mell_{\loc}$ and $\kappa>e^{-m'\ell}$ with $L^{2d}\leq e^{m'\ell}$ the following holds. If $\omega\in\Omega_{\kappa}^{g}$ and $I\subset (-\infty,E_{\loc}]$ an interval with $|I|\leq \kappa$, then
\begin{enumerate}
\item[(i)] there exists $x=x_{\omega}\in\Lambda_L$ such that
$\tr \id_I(H_{\omega,L}) \chi_y\le e^{-m'\ell}$ for all $y\in\Lambda_L$ such that $|x-y| >3\ell+8R =:  \ell'$,
\item[(ii)] 
$\tr \id_I(H_{\omega,L}) \leq C_1' \ell^d$, with constant $C'_1$ specified in \eqref{lem1step3}.
\end{enumerate}
\end{lem}

\begin{proof}[Proof of Lemma \ref{cont:lem1}]

Let $I$ and $\omega$ as in the lemma's statement and let $(\psi_i,\lambda_i)_{i\in\mathcal{I}}$ be the collection of eigenpairs of $ H_{\omega,L}$ with $\lambda_i\in I$. For now we denote the localization centers of $\psi_i$, i.e. the points specified by Lemma \ref{cont:lem5}, by $x_{i}$. Since $\omega\in\Omega_{\kappa}^{\W}$ we thus have $\dist\big(\sigma\big( H_{\omega,\Lambda^L_{2\ell+4R}(z)}\big),I\big)>\kappa$ for all $z\in\Lambda_L$ with $|z-x_1|\geq 2\ell+6R$. Since by assumption $\kappa>e^{-m'\ell}$ this implies that $|x_i-x_1| < 2\ell+6R$ for all $i\in\mathcal I$.
For the first statement let $x:=x_1$. Because $|\mathcal I | = \tr\id_I(H_{\omega,L}) \leq C_1 L^d$ with $C_1$ as in Lemma \ref{lem:AprioriBound} it follows that 
\beq
\label{lem1step2}
 \tr \id_I( H_{\omega,L})\chi_y \leq C_1 L^d e^{-m'\ell} 
\eeq 
for all $y\in\Lambda_L$ such that $|x-y| >3\ell+8R$. Because $L^{2d} \leq e^{m'\ell}$ this proves $(i)$.
For the second assertion, we use that
\begin{align}
\tr{\chi_{{\Lambda_L\setminus\Lambda^L_{6\ell+16R}(x)}}\id_I( H_{\omega,L})} &\leq \sum_{\substack{y\in\Z^d:\\|y-x|>3\ell+8R}}\tr{\id_I(H_{\omega,L})\chi_y}\notag\\
&\le   C_1 L^{2d}e^{-m'\ell} \leq C_1
\label{pf:cont:lem1eq2}
\end{align}
holds by \eqref{lem1step2}. This gives the estimate
\begin{align}
\tr \id_I( H_{\omega,L})&\le C_1+\tr\chi_{{\Lambda^L_{6\ell+16R}(x)}}\id_I( H_{\omega,L})\le C_1 (1+(6\ell+16R)^d)\notag\\
&\leq C_1' \ell^d. \label{lem1step3}
\end{align}
\end{proof}

{\begin{proof}[Proof of Theorem \ref{thm:MainResult}]
The proof is similar to the one of Theorem \ref{thm:MainResultv2}. First, let $L\geq \ell \geq \mell_{\loc}$ and $\min\{E_{\loc},E_{\spc}\}>\kappa>0$ such that $\kappa>e^{-m'\ell}$ and $L^{2d}\leq e^{m'\ell}$ hold. We again start by choosing a fixed $E\in (0,\min\{E_{\loc},E_{\spc}\})$ and decompose the interval $[0,E]$ into a family $(K_i)_{i\in\mathcal I}$ of intervals with side length $|K_i| = \kappa$, with $|K_{i+1}\cap K_i| \geq \kappa/2$ and such that $|\mathcal I| \leq 4 E_{\spc} \kappa^{-1} + 1$. We also set $K_{i,8\veps} := K_i + [-8\veps,8\veps]$ for $\veps \in (0,1/12)$. By Wegner's estimate
\beq
\label{pf:ProbEstBuffer}
\P\big(\tr \id_{K_{i,8\veps}\setminus K_i}( H_{\omega,L}) = 0 ) \geq 1- 16 C_{\W} L^d \veps. 
\eeq
If we define the event
\beq
\label{pf:conteq1}
\Omega_{i,\kappa}^g := \Omega_{\kappa}^g\cap \big\{\tr \id_{K_{i,8\veps}\setminus K_i}( H_{\omega,L}) = 0 \big\},
\eeq
then for $0<\delta<\kappa/2$ we obtain from \eqref{pf:ProbEstBuffer} and \eqref{eq:ProbEstGoodSet} the bound
\begin{align}
\label{pf:conteq2}
&\P\pa{\spac_{E}(H_{\omega,L})<\delta} \notag\\
&\quad\leq \P\big(\big\{\spac_{E}(H_{\omega,L})<\delta\big\} \cap \Omega^{g}_\kappa\big) + 6 C^{\prime 2}_{\W} L^{2d} \ell^{2d} \kappa+ L^{2d} e^{-m'\ell}\notag\\
&\quad\leq \sum_{i\in\mathcal{I}} \P\big(\set{\spac_{K_i}(H_{\omega,L})<\delta}\cap \Omega_{i,\kappa}^g\big) + \Xi_{L,\ell,\kappa,\veps}.
\end{align}
Here we also abbreviated $\Xi_{L,\ell,\kappa,\veps}:= C''_{\W} L^d \kappa^{-1} \veps + C''_{\W} L^{2d}\ell^{2d} \kappa+ L^{2d} e^{-m'\ell}$ for a suitable constant $C''_{\W}$. Lemma \ref{cont:lem1} implies that for fixed $i\in\mathcal I$ and $\omega\in \Omega_{i,\kappa}^g$ there exists $x_{i,\omega}\in\Lambda_L$ (which we can assume without loss of generality is in $\Lambda_L^{\#}:= \Lambda_{L}\cap\Z^d$) such that $P_{i,\omega} := \id_{K_{i,\veps}}(H_{\omega,L})$ is localized with localization center $x_{i,\omega}$:
\beq
\tr \chi_x P_{i,\omega}  \leq  e^{-m'\ell} \leq e^{-m''\ell'}
\eeq
for all $x\in \Lambda_L$ with $|x-x_{i,\omega}| \geq 3\ell+8R = \ell'$ and a suitable $0<m''<m'$. If we define 
\begin{align}
\Omega_{i,x}^{\loc} &:= \{P_{i,\omega} \text{ is localized with localization center } x\},\\
\Omega_{i}^{\spa} &:= \big\{\spac_{K_{i,\veps}}(H_{\omega,L}) < \delta\big\},
\end{align}
then we arrive at
\beq
\eqref{pf:conteq2} \leq \sum_{i\in\mathcal I} \sum_{x\in \Lambda_L^{\#}}  \P\pa{ \Omega_{i}^{\spc} \cap \Omega_{i,x}^{\loc} \cap \Omega_{i,\kappa}^g} + \Xi_{L,\ell,\kappa,\veps}.
\eeq
Next we again partition the configuration space into subcubes, but now only in a spacial neighbourhood of the localization center $x$. More precisely, we partition $[0,1]^{\Gamma_{\ell',x}}$ into (not necessarily disjoint) cubes $Q_{j,x}\subset [0,1]^{\Gamma_{\ell',x}}$, $j\in\mathcal{J}$, of side length $2\veps$ and such that
\beq
|\mathcal{J}| \leq ((2\veps)^{-1}+1)^{|\Gamma_{\ell',x}|} \quad \text{ and } \quad \sum_{j\in\mathcal J}\P(Q_j) \leq 1+ 4 \veps |\Gamma_{\ell',x}|  \rho_+.
\eeq 
We denote the centers of $Q_{j,x}$ by $\omega_{0,j,x}\in [0,1]^{\Gamma_{\ell',x}}$, i.e. $Q_{j,x} = \omega_{0,j,x}+[-\veps,\veps]^{\Gamma_{\ell',x}}$ . So far we estimated
\begin{align}
{\eqref{pf:conteq2}} &\leq \sum_{i\in\mathcal{I}}\sum_{x\in\Lambda_L^\#}\sum_{j\in\mathcal{J}} \P\pa{\big(Q_{j,x}\times [0,1]^{\Gamma_L\setminus\Gamma_{\ell',x}}\big) \cap  \Omega_{i}^{\spc}  \cap \Omega_{i,\kappa}^g \cap \Omega_{i,x}^{\loc}} + \Xi_{L,\ell,\kappa,\veps}.
\label{PfSpac2}
\end{align}
Let $i\in\mathcal{I}$, $x\in \Lambda_L^\#$ and $j\in\mathcal{J}$ be fixed and such that the probability on the right hand side of \eqref{PfSpac2} is non-zero. For a set $A\subset [0,1]^{\Gamma_L}$ let 
\beq
\proj_{\Lambda^c_{\ell'}(x)}^{Q_{j,x}}(A) := \big\{\omega|_{\Lambda_{\ell'}^c(x)}:\, \omega\in A \text{ and } \omega|_{\Lambda_{\ell'}(x)}\in Q_{j,x}\big\} \subseteq[0,1]^{\Gamma_L\setminus\Gamma_{\ell',x}}.
\eeq
We now estimate the probability in \eqref{PfSpac2} by
\beq
\label{PfSpac3}
\P\pa{\pa{Q_{j,x}\times  \proj_{\Lambda^c_{\ell'}(x)}^{Q_{j,x}}\big( \Omega_{i}^g\cap\Omega_{i,x}^{\loc}\big)} \cap \Omega_{i}^{\spc}}
\eeq
and choose a fixed  
\beq
\label{pfSpac11}
\omega_{1,\Lambda_{\ell'}^c(x)} \in \proj_{\Lambda^c_{\ell'}(x)}^{Q_{j,x}}\big( \Omega_{i}^g\cap\Omega_{i,x}^{\loc}\big) \neq \emptyset. 
\eeq
Here the dependence on $i$ and $j$ is suppressed in notation. By construction, there exists $\omega_{1,\Lambda_{\ell'}(x)}\in Q_{j,x}$ such that $\omega_1:= (\omega_{1,\Lambda_{\ell'}(x)},\omega_{1,\Lambda_{\ell'}^c(x)})\in \Omega_i^g \cap \Omega_{i,x}^{\loc}$, where also the dependence on $x$ is suppressed in notation.
Hence, Lemma \ref{lem:GoodConf1} can be applied for $\ell'$ as small scale, $m''$ as inverse localization length in \eqref{eq:locP}, $n\leq C'_1\ell^d$ and $r=d+1$. This yields a configuration $\what\omega\in Q_{\veps}^{(x,\ell')}(\omega_1,\omega_{0,j})$ such that 
\beq
\spac_{I_{i,\veps}}(H_{\widehat\omega,L})\ge 8\veps \ell^{\prime -\ell^{\prime d}2 C'_1(2d+2r)}.
\eeq
Lemma \ref{cont:dimlem3} is now applicable for $n\leq C'_1\ell^d$, $\delta_0 = 8\veps \ell^{\prime-\ell^{\prime d}2C'_1(2d+2r)}$ and the family $(\omega_j)_{j\in\Gamma_{\ell',x}}$ of random variables. This yields
\begin{align}
&\big|\big\{\omega\in Q_{\veps}^{(x,\ell')}(\omega_{1},\omega_{0,j}):\, \spac_{K_{i,\veps}}(H_{\omega,L}) < \delta\big\}\big|_{\Lambda_{\ell'}(x)} \notag\\
&\qquad\leq  c'_1 \ell^{\prime d} (2\veps)^{|\Gamma_{x,\ell'}|} \exp\pa{ \frac{-c'_2|\log \delta|}{\ell^{\prime 2d}(|\log \veps|+\ell^{\prime d+1})}}.
\end{align}
Here $|A|_{\Lambda_{\ell'}(x)}$ stands for the $|\Gamma_{\ell',x}|$-dimensional Lebesgue measure of a set $A$.
Because this bound is independent of the $\omega_{1,\Lambda_{\ell'}^c(x)}$ chosen in \eqref{pfSpac11}, we can use \eqref{PfSpac3} to estimate 
\begin{align}
&\P\pa{\big(Q_{j,x}\times [0,1]^{\Gamma_L\setminus\Gamma_{\ell',x}}\big) \cap  \Omega_{i}^{\spc}  \cap \Omega_{i,\kappa}^g \cap \Omega_{i,x}^{\loc}} \notag\\
&\qquad \leq c_1'\ell^{\prime d} (2\veps)^{|\Gamma_{\ell',x}|}   \exp\pa{\ell^{\prime d}\log \rho_+ - \frac{c'_2|\log \delta|}{\ell^{\prime 2d}(|\log \veps|+\ell^{\prime d+1})}}.
\end{align}
Overall, we arrive at 
\begin{align}
\eqref{PfSpac2} &\leq  c''_1 L^{d}\kappa^{-1}  \exp\pa{\ell^{\prime d}\log \rho_+ - \frac{c'_2|\log \delta|}{\ell^{\prime 2d}(|\log \veps|+\ell^{\prime d+1})}}\notag\\
&\quad+  C''_{\W} L^d \kappa^{-1} \veps + C''_{\W} L^{2d}\ell^{2d} \kappa+ L^{2d} e^{-m'\ell}. \label{decay1}
\end{align}
We now choose $\veps := \exp \pa{-|\log \delta|^{1/4}}$, $\kappa := \exp\pa{-|\log \delta|^{1/8}}$ and $\ell = |\log\delta|^{1/(8d)}$, which yields
\begin{align}
\P\pa{\spac_{E}(H_{\omega,L})<\delta} &\leq C'_{\spc} L^{2d}\pa{ e^{-m''|\log\delta|^{1/(8d)}} + e^{|\log\delta|^{1/8}(1+\rho_+)-c_2'|\log\delta|^{1/2}}} \notag\\
&\leq C_{\spc} L^{2d}  e^{-|\log\delta|^{1/(9d)}},
\label{decay2}
\end{align}
for $\delta\leq \delta_0$, where $\delta_0>0$ is sufficiently small. Finally, the condition $\kappa>e^{-m'\ell}$ is satisfied for sufficiently large $L$ and the conditions $L\geq \ell$ and $L^{2d}\leq e^{m'\ell}$ are satisfied for
\beq
\label{decay3}
\exp\big(-L^{8d}\big) \leq \delta \leq \exp\big(-(\log L)^{9d}\big).
\eeq
If $\delta < \exp(-L^{8d})$ we can omit the introduction of a second scale $\ell\ll L$ and directly carry out the argument on the whole box $\Lambda_L$, in a similar fashion as in the proof of Theorem \ref{thm:MainResultv2}.
\end{proof}

}

\section{Proof of the Minami-type estimate}
\label{sec3}

Before we start with the proof of Theorem \ref{cor2} we make some preliminary remarks. Let $H^\mu_\omega=-\mu\Delta+V_\omega$ be the standard random Schr\"odinger operator from Section \ref{sec1}. The random operator
\beq
\label{def:auxOp}
\wt H^{\mu,E}_{\omega} := V^{-1/2} (H_{\omega}-E) V^{-1/2} = -\mu V^{-1/2}\Delta V^{-1/2} + \wt V^E_o + \wt V_\omega
\eeq
is a deformed random Schr\"odinger operator with periodic potential $\wt V^E_o := -EV^{-1}$ and random potential $\wt V_\omega := \sum_{k\in\Z^d} \omega_k \wt V_k$, where $\wt V_k := V^{-1} V_k$. We stress the dependence on $\mu$ in notation because, as mentioned earlier, we'll have to work with $L$-dependent couplings $\mu_L$ in some small neighbourhood of a fixed $\mu$. 

Tracking constants in Section \ref{sec:SpacingNoLoc} shows the following. For fixed $E_0\in(0,E_{\M})$, with $E_{\M}$ as defined in \eqref{def:EMin}, and $K>0$ there exists $\veps>0$ and constants  $\mell_{\spc},C_{\spc}>0$ such that for all $\mu' \in [\mu-\veps,\mu+\veps]$ and all $E\in[0,E_0]$
\beq
\label{eq:PrepThm}
\P\pa{\spac_{[-\veps,\veps]}(\wt H^{\mu',E}_{\omega,L}) < \delta} \leq C_{\spc} L^{2d} |\log\delta|^{-K}
\eeq
holds for all $L\geq \mell_{\spc}$ and $\delta <1$.

\begin{proof}[Proof of Theorem \ref{cor2}]
For fixed $E_0\in(0,E_{\M})$ and $K>0$ we denote by $\veps,\mell_{\spc},C_{\spc}$ the constants from above. After possibly enlarging $\mell_{\spc}$ we have $\delta \leq \mell_{\spc}^{-d}\leq \veps/2$ and $4 \delta L^d \leq 1$ for $L,\delta$ which satisfy $L \geq \mell_{\spc}$ and $\delta \leq \exp(-(\log L)^{5d})$. 

Let now $E\in[0,E_0]$, $L\geq \mell_{\spc}$ and $0<\delta\leq \exp\pa{-(\log L)^{5d}}$ be fixed. Our startint point is Lemma \ref{lem:SpecCong}, which, applied for $A=H^{\mu}_{\omega,L}-E$, $S=V_-^{1/2} V^{-1/2}$ and $\veps=\delta V_-/2$, yields 
\begin{align}
\label{anothertry1}
\tr \id_{[E-\delta V_-, E+ \delta V_-]}(H^{\mu}_{\omega,L}) &= \tr \id_{[-\delta V_-, \delta V_-]}(H^{\mu}_{\omega,L}-E)\notag\\
&\leq \tr \id_{[-\delta, \delta]}(\widetilde H_{\omega,L}^{\mu,E}).
\end{align}
By $\widetilde E^{\mu ,E}_{\omega,j}, j\in \N$, we denote the eigenvalues of $\widetilde H_{\omega,L}^{\mu,E}$ in ascending order. If $C_1$ denotes the constant from Lemma \ref{lem:AprioriBound}, then
\beq
\label{anothertry2}
\P\pa{\tr \id_{[-\delta, \delta]}(\widetilde H_{\omega,L}^{\mu,E}) \geq 2} \leq \sum_{j=1}^{C_1L^d} \P\pa{\spac_{[-\veps/2,\veps/2]}(\widetilde H_{\omega,L}^{\mu,E})<2\delta,\, \widetilde E^{\mu ,E}_{\omega,j}\in[-\delta,\delta]},
\eeq
where we used that $\delta\leq \veps/2$.
In the sequel each term on the right hand side is estimated separately. Let's first introduce some notation. Let $N\in\N$ such that $(2L^d\delta)^{-1}-1 < N \leq (2L^d\delta)^{-1}$ and 
\beq
\label{anothertry4}
I_i:= [-\delta,\delta] + (i-1)2\delta \quad \text{ for } i\in\set{1,...,N}.
\eeq
Moreover, for $i\in \set{1,...,N}$, $j\in\N$ and $\theta>0$ we define
\beq
\label{anothertry5}
\Omega_{i,j}^{\theta} := \big\{\spac_{[-\theta,\theta]}(\widetilde H^{\mu,E}_{\omega,L}) < 2\delta\big\} \cap \big\{\widetilde E^{\mu ,E}_{\omega,j} \in I_i\big\}. 
\eeq
Let $\kappa:=(1+L^{-d})^{-1}$. Then we claim that for some constant $C_\rho$, that only depends on the single-site density $\rho$,
\begin{align}
\P\pa{\Omega_{1,j}^{\veps/2}} \leq C_\rho\, \P\pa{\spac_{[-\veps,\veps]}(\widetilde H^{\kappa\mu,\kappa E}_{\omega,L}) < 2\delta,\, \widetilde E^{\kappa\mu,\kappa E}_{\omega,j} \in \kappa I_i}.\label{anothertry6}
\end{align}
In this case, summation of \eqref{anothertry6} over $i\in\set{1,...,N}$ yields
\beq
\label{anothertry7}
\P\pa{\Omega_{1,j}^{\veps/2}} 4 \leq C_\rho L^d \delta\, \P\pa{\spac_{[-\veps,\veps]}(\widetilde H^{\kappa\mu,\kappa E}_{\omega,L}) < 2\delta},
\eeq
where we used that $N^{-1} \leq 4 L^d\delta$ and that for $i_1\neq i_2$
\beq
\label{anothertry8}
\big\{\widetilde E^{\kappa\mu,\kappa E}_{\omega,j} \in \kappa I_{i_1}\big\} \cap \big\{\widetilde E^{\kappa\mu,\kappa E}_{\omega,j} \in \kappa I_{i_2}\big\} = \emptyset.
\eeq
The statement now follows from an application of \eqref{eq:PrepThm} to the right hand side of \eqref{anothertry7}.

We are left with proving \eqref{anothertry6}. For the operator $\wt H_{\omega,L}^{\mu,E}$ a shift of random couplings results in an energy shift. If we denote ${\bm\tau} = (\tau,...,\tau)\in \Gamma_L$ for fixed $\tau\in\R$, then
\beq
\label{pf:LocMin1}
\wt H_{\omega+{\bm \tau},L}^{\mu,E} = \wt H_{\omega,L}^{\mu,E} + \tau  \chi_{\Lambda_L} V V^{-1}\chi_{\Lambda_L} = \widetilde H_{\omega,L}^{\mu,E} + \tau 
\eeq
as operators on $L^2(\Lambda_L)$. This implies that 
\beq
\label{pf:LocMin2}
\spac_{K}(\wt H_{\omega,L}^{\mu,E}) = \spac_{K+\tau}(\widetilde H_{\omega+{\bm \tau},L}^{\mu,E})
\eeq
for any interval $K\subset\R$. Let $\eta_i := (i-1)2\delta$ denote the centers of the intervals $I_i$. The change of variables $\omega_k \to \omega_k+\eta_i$ and \eqref{pf:LocMin2} give
\begin{align}
\P\pa{\Omega_{1,j}^{\veps/2}}  &\leq \int_{[\eta_i,1+\eta_i]^{\Gamma_L}} \id_{\Omega_{i,j}^\veps}(\omega) \prod_{k\in\Gamma_L} \rho(\omega_k-\eta_i) d\omega_k,
\label{pf:LocMin4}
\end{align}
where we also used $\eta_i \leq L^{-d} \leq \veps/2$ and \eqref{pf:LocMin2}.
Another change of variables $\omega_k \to \kappa\omega_k$ yields
\beq 
\label{pf:LocMin5}
\eqref{pf:LocMin4} \leq \kappa^{-|\Gamma_L|} \int_{[a_{i},b_{i}]^{\Gamma_L}} \id_{\Omega_{i,j}^{\veps}}(\kappa^{-1}\omega) \prod_{k\in\Gamma_L} \rho(\kappa^{-1}\omega_k-\eta_i) d\omega_k,
\eeq 
where $a_{i} := \kappa\eta_i$ and $b_{i} := \kappa(1+\eta_i)$ (which both depend on $L$ through $\kappa$). Note that we have
\beq
\label{pf:LocMin6}
\widetilde H^{\mu,E}_{\kappa^{-1}\omega,L} = \kappa^{-1} \widetilde H^{\kappa\mu,\kappa E}_{\omega,L},
\eeq
and hence by definition of the events $\Omega_{i,j}^\veps$
\beq
\label{pf:LocMin7}
\kappa^{-1}\omega\in \Omega_{i,j}^\veps \iff \omega\in \kappa\Omega_{i,j}^\veps \iff
\begin{cases} 
\spac_{\kappa\veps}(\widetilde H^{\kappa\mu,\kappa E}_{\omega,L}) < \kappa 2 \delta \\
\qquad\text{and} \\ 
\widetilde E^{\kappa \mu,\kappa E}_{\omega,j} \in \kappa I_i.
\end{cases}
\eeq
Because $\kappa <1$ the relation \eqref{pf:LocMin7} yields
\beq
\label{anothertry9}
\kappa\Omega_{i,j}^\veps \subset \big\{\spac_{\veps}(\widetilde H^{\kappa\mu,\kappa E}_{\omega,L}) < 2 \delta,\, \widetilde E^{\kappa \mu,\kappa E}_{\omega,j} \in \kappa I_i\big\}.
\eeq 
Moreover, since $\rho$ satisfies $(V_4)$ we have for $x\in (a_{i},b_{i})\subset(0,1)$ that $\kappa^{-1}x-\eta_i \in (0,1)$ as well and
\beq
\label{pf:LocMin8}
\rho(\kappa^{-1}x-\eta_i) \leq \rho(x) +2 \mathcal K L^{-d} \leq \rho(x) \pa{1+\frac{2\mathcal K}{L^d\rho_-}}
\eeq
Estimating \eqref{pf:LocMin5} via \eqref{anothertry9} and \eqref{pf:LocMin8} yields
\beq
\label{pf:LocMin9}
\eqref{pf:LocMin5} \leq C_\rho\, \P\pa{\spac_{\veps}(\widetilde H^{\kappa\mu,\kappa E}_{\omega,L}) < 2 \delta,\, \widetilde E^{\kappa \mu,\kappa E}_{\omega,j} \in \kappa I_i}.
\eeq
\end{proof}

\section{Simplicity of spectrum and Poisson statistics}
\label{sec4}

As mentioned in Section \ref{sec1}, both statements follow from Theorem \ref{thm:MainResult} respectively Theorem \ref{cor2} and the techniques from \cite{KM,CGK} respectively \cite{Min,Molchanov,CGK}. For convenience we recap the arguments here, closely sticking to the above references.

For the proof of Corollary \ref{cor3} we apply the following consequence of \eqref{def:LocFMB}: With probability $1$, for any normalized eigenpair $(\psi,\lambda)$ of $H_\omega$ with $\lambda<E_{\loc}$ there exists a constant $C_\psi$  such that for all $x\in\R^d$
\beq
\label{def:loc2}
\|\psi\|_x \leq C_\psi e^{-m|x|}.
\eeq
Here, the localization center has been absorbed into the ($\omega$-dependent) constant $C_\psi$.

\begin{proof}[Proof of Corollary \ref{cor3}]
Let $E<\min\{E_{\spc},E_{\loc}\}$ be fixed. First we note that by Theorem \ref{thm:MainResultv2} there exists $\mell_0$ such that for $L\geq \mell_0$
\beq
\label{simplSpec1}
\P\pa{\spac_{E}(H_{\omega,L}) < 3e^{-\sqrt{L}}}\leq L^{-2}.
\eeq
Since the right hand side is summable over $L\in \N$ the Borel-Cantelli lemma yields that the set
\beq
\label{simplSpec2}
\Omega_\infty:= \big\{ \spac_{E}(H_{\omega,L}) < 3 e^{-\sqrt{L}} \text{ for infinitely many }L\in\N\big\}
\eeq
is of measure zero with respect to $\P$. Let $\Omega_{\loc}$ be the set of measure one such that \eqref{def:loc2} holds for all $\omega\in\Omega_{\loc}$. We now choose a fixed 
\beq
\omega\in \Omega_{\loc} \cap \set{\exists E' \leq E:\, \tr \id_{\set{E'}}(H_\omega) \geq 2} =: \Omega_{\loc} \cap \Omega_{\geq 2};
\eeq 
i.e. for the configuration $\omega$ there exists $E'\leq E$ such that $E'$ is an eigenvalue of $H_\omega$ with two linearly independent, normalized and exponentially decaying eigenfunctions $\phi,\psi$. We now apply \cite[Lemma 1]{KM} with the slightly modified choice $\veps_L = L^d e^{-m L/2} \ll e^{-\sqrt L}$. The lemma is formulated for the lattice but generalizes to the continuum as has been remarked in \cite{CGK}. This implies that for $I_L:= [E-e^{-\sqrt L},E+e^{-\sqrt L }]$ and all sufficiently large $L\in \N$
\begin{equation}
\label{simplSpec3}
\tr \id_{I_L}(H_{\omega,L}) \geq 2
\end{equation}
holds, and consequently $\Omega_{\loc} \cap \Omega_{\geq 2} \subset \Omega_\infty$. The latter set is of $\P$-measure zero, and the result follows from $\P\pa{\Omega_{\loc} \cap \Omega_{\geq 2}}=0$.
\end{proof}

\begin{proof}[Proof of Theorem \ref{cor4}]
The proof closely follows the one in \cite[Section 6]{CGK}. Let $E\in [0,\min\{E_{\M},E_{\loc}\}]$ be fixed and such that $n(E)>0$. The starting point is to construct a triangular array of point processes which approximate $\xi_{\omega}^L:=\xi_{E,\omega}^L$ sufficiently well. To this end, let $L$ be fixed and $\ell:= (\log L)^2$. Then we define point processes $\xi_{\omega}^{L,m}$ for $m\in \Upsilon_{L} := (\ell+2\lceil R \rceil) \Z^d\cap \Lambda_{L-\ell}$ via $\xi^{L,m}_{\omega}(B):= \tr \id_{E+L^{-d}B}(H_{\omega,\Lambda_\ell(m)})$ ($B\subset\R$ Borel measurable). This definition ensures that for $m,n\in \Upsilon_{L}$, $m\neq n$, the processes $\xi^{L,m}_{\omega}$ and $\xi^{L,n}_{\omega}$ are independent. 

The proof now consists of two parts. In the first part one shows that the superposition $\wt \xi_{\omega}^{L}:=\sum_{m\in\Upsilon_{L}}\xi^{L,m}_{\omega}$ is a good approximation of the process $\xi_{\omega}^{L}$ in the sense that, if one of them converges weakly, then they share the same weak limit. This is a consequence of spectral localization, and the arguments are very similar to \cite{CGK}. However, slight adaptions are in place since we work with different finite-volume restrictions of $H_\omega$. We comment on this below. In the second part one then proves that the process $\wt \xi_{\omega}^{L}$ weakly converges towards the Poisson point process with intensity measure $n(E) dx$. This is the case if and only if for all bounded intervals $I\subset\R$ the three properties
\begin{align}
\lim_{L\to\infty} \max_{m\in \Upsilon_{L}} \P\pa{ \xi_{\omega}^{L,m}(I)\geq 1 } = 0, \label{PoissonA1}\\
\lim_{L\to\infty} \sum_{m\in\Upsilon_{L}} \P \pa{\xi_\omega^{L,m} \geq 1} = |I|n(E),\label{PoissonA2}\\
\lim_{L\to\infty} \sum_{m\in\Upsilon_{L}} \P\pa{ \xi_{\omega}^{L,m}(I) \geq 2 } = 0\label{PoissonA3}
\end{align}
hold. We assume for convenience that $|I|\leq 1$ and note that \eqref{PoissonA1} follows from Wegner's estimate. Let $L$ be sufficiently large such that $\ell\geq \mell_{\M}$, where $\mell_{\M}$ is the initial scale from Theorem \ref{cor2}. We can then apply the Theorem for $K=12d$ to estimate 
\beq
\label{Poisson1}
 \P\pa{ \xi_{\omega}^{L,m}(I) \geq 2 } \leq C'_{\M} \ell^{-2d} L^{-d} 
\eeq
for all $m\in \Upsilon_{L}$, which ensures \eqref{PoissonA3}. Moreover, for $n > C_1\ell^d$ (with $C_1$ as in Lemma \ref{lem:AprioriBound}) we have $\P \big(\xi_\omega^{L,m} \geq n\big) = 0$. The estimate
\begin{align}
\sum_{m\in\Upsilon_{L}} \sum_{n=2}^\infty  \, \P \pa{\xi_\omega^{L,m}(I) \geq n} &\leq C_1 \ell^d |\Upsilon_{L}|  \sup_{m\in\Upsilon_{L}} \P \pa{\xi_\omega^{L,m}(I) \geq 2}\notag \\
&\leq C''_{\M} \ell^{-d}  \label{Poisson2}
\end{align}
then readily yields \eqref{Poisson3}. Moreover, it also shows that \eqref{PoissonA2} would follow from
\beq\label{Poisson3}
\lim_{L\to\infty} \sum_{m\in\Upsilon_{L}} \E\left[ \xi_{\omega}^{L,m}(I) \right] = n(E)|I|.
\eeq
To verify \eqref{Poisson3}, we will use the following lemma, which is a slight variant of \cite[Lemma 6.1]{CGK}.

\begin{lem}\label{lem:Trace}
For bounded intervals $J\subset\R$ we have
\begin{align}
 \lim_{L\to\infty} \E \big[\big|\wt\xi_{\omega}^{L}(J)- \xi_{\omega}^{L}(J)\big|\big] = 0,\label{eq:TraceLem1}\\
 \lim_{L\to\infty} \E \big[\big|\Theta_\omega^L  - \xi_{\omega}^{L}(J)\big|\big] = 0,\label{eq:TraceLem2}
\end{align}
where $\Theta_\omega^L(J) := \tr \chi_{\Lambda_L}\id_{E+L^{-d}J}(H_{\omega})$.
\end{lem}

A sketch of proof for the lemma is given below. By combining \eqref{eq:TraceLem1} and \eqref{eq:TraceLem2} we obtain
\beq
\lim_{L\to\infty} \sum_{m\in\Upsilon_{L}} \E\left[ \xi_{\omega}^{L,m}(I) \right] = \lim_{L\to\infty} \E \left[ \Theta_\omega^L \right ] = n(E) |I|
\eeq
for the interval $I$ from above.
Hence \eqref{Poisson1}--\eqref{Poisson3} hold and $ \wt\xi_{\omega}^{L}$ converges weakly to the Poisson process with intensity measure $n(E)dx$. As argued in \cite{CGK}, the convergence \eqref{eq:TraceLem1} and the density of step functions in $L^1$ is sufficient to prove that $\xi_\omega^L$ weakly converges to the same limit as $\wt \xi_{\omega}^{L}$. 
\end{proof}

\begin{proof}[Proof of Lemma \ref{lem:Trace}]
We first note that for our model a local Wegner estimate holds, i.e. there exists $C'_{\W}$ such that
\beq\label{property1}
\sup_{x\in \R^d\cap\Lambda_L}\E \left[ \chi_x \id_{J}(H_{\omega,L}) \right] \leq C'_{\W} |J|
\eeq
for all intervals $J\subset (-\infty,E_{\M}]$. This is proved in \cite[Theorem 2.4]{CGK1} for periodic boundary conditions, but the argument also applies for Dirichlet boundary conditions. 
The second ingredient of the proof is the following consequence of spectral localization \cite[Theorem 3.2]{DGM}. There exist constants constants $C'_{\loc},m'>0$ such that the following holds: For open sets $G\subset G'\subset\R^d$ with $\dist(\partial G',\partial G) \geq 1$ and $a\in G$ we have
\beq\label{property2}
\E \left[ \norm{\chi_a\pa{\id_{J}(H_{\omega,G})-\id_{J}(H_{\omega,G'})}\chi_a}_1 \right] \leq C'_{\loc} e^{-m'\dist(a,\partial G)}
\eeq
for all intervals $J\subset (-\infty,E_{\M}]$. 
We now establish \eqref{eq:TraceLem1}. The proof of \eqref{eq:TraceLem2} is similar.  To this end, we split each $\Lambda_\ell(m)$, $m\in\Upsilon_{L}$, into a bulk part $\Lambda_\ell^{(i)}(m) :=  \Lambda_{\ell-\ell^{2/3}}(m)$ and a boundary part $\Lambda_\ell^{(o)}(m) := \Lambda_{\ell}(m)\setminus\Lambda_{\ell}^{(i)}(m)$. If we abbreviate $J_{E,L}:= E+L^{-d} J$ then this splitting yields
\begin{align}
\E \left[\big|\wt\xi_{\omega}^{L}(J)- \xi_{\omega}^{m,L}(J)\big|\right] =& \sum_{m\in\Upsilon_{L}} \E\left[ \big|\tr \chi_{\Lambda^{(i)}_\ell(m)}\pa{\id_{J_{E,L}}(H_{\omega,\Lambda_\ell(m)}) - \id_{J_{E,L}}(H_{\omega,L})}\big|\right]\notag\\
&+\sum_{m\in\Upsilon_{L}} \E\left[ \big|\tr \chi_{\Lambda^{(o)}_\ell(m)}\pa{\id_{J_{E,L}}(H_{\omega,\Lambda_\ell(m)}) - \id_{J_{E,L}}(H_{\omega,L})}\big|\right]\notag\\
&\quad+ \E \Big[\tr \Big(\chi_{\Lambda_L}-\sum_{m\in\Upsilon_{L}}\chi_{\Lambda_{\ell}(m)}\Big)\id_{J_{E,L}}(H_{\omega,L})\Big]\notag\\
&=: (\text{bulk}) + (\text{boundary}) + (\text{rest}).
\end{align}
For the latter two terms we apply the local Wegner estimate from \eqref{property1} to get
\begin{align}
(\text{boundary}) &\leq |\Upsilon_{L}| C'_{\W} L^{-d} d \ell^{d-1}(\sqrt\ell+2R)  \leq C''_{\W}\ell^{-1/2},\label{Poissonsmallterm1}\\
(\text{rest}) &\leq  C'_{\W} L^{-d} |\Upsilon_L| \ell^{d-1} (2R+2) \leq C_{\W}'''\ell^{-1}\label{Poissonsmallterm2}.
\end{align}
On the bulk contribution we in turn apply localization via \eqref{property2} to get 
\beq
(\text{bulk}) \leq |\Upsilon_{L}| C'_{\loc} \ell^{d} e^{-m' \ell^{2/3}} = C''_{\loc} L^d e^{-m'\ell^{3/2}}.\label{Poissonsmallterm3}
\eeq
Because $L=e^{\sqrt\ell}$ all three terms \eqref{Poissonsmallterm1}--\eqref{Poissonsmallterm3} converge to zero as $L\to\infty$. 
\end{proof}

\begin{appendix}

\section{Properties of deformed Schr\"odinger operators}
\label{sec:appB}

In this appendix we consider random deformed operators $H_\omega:= -\mu G\Delta G + V_o + V_\omega$. The assumptions on $G,V_o$ and $V_\omega$ are the same as in Section \ref{sec:SpacingDeformedOp}. The Lemmas \ref{lem:AprioriBound} and \ref{lem:WegnerDef} below establish two technical properties of deformed RSO which enter the proof of Theorem \ref{thm:MainResultv2}, an a priori trace bound and Wegner's estimate.

Both of them are proven by rewriting the respective estimates in terms of a standard RSO via the following lemma. 

\begin{lem}\label{lem:SpecCong}
Let $A$ be a self-adjoint operator on a separable Hilbert space $\mathcal H$, let $S$ be an invertible contraction on  $\mathcal H$ (i.e. $\|S\|\le1$), and let $C_\veps(A):=\tr \id_{[-\veps,\veps]}(A)$. Then we have 
\beq
C_\veps(A)\le C_\veps(SAS^*).
\eeq
\end{lem}
\begin{proof}
Consider $B:= \id_{\R\setminus [-\veps,\veps]}(A)A$. Then $C_0(B)=C_\veps(A)$ and, by Sylvester's law of inertia, we have $C_0(SBS^*)=C_0(B)$. But
\[SAS^*=SBS^*+S\id_{[-\veps,\veps]}(A)AS^* \quad\text{ and }\quad \norm{S\id_{[-\veps,\veps]}(A)AS^*}\le\veps,\]
so Weyl's inequality implies that
\beq
C_0(SBS^{*}) \leq C_\veps(SAS^*).
\eeq
\end{proof}

\begin{lem}[A priori bound]
\label{lem:AprioriBound}
For every $E<\infty$ we have for (almost) every $\omega$ and $L>0$
\beq\label{eq:aprsp}
\tr{\id_{(-\infty,E]}\pa{H_{\omega,L}}}\le C_EL^d.
\eeq
\end{lem}
\begin{proof}
With the constant $c:= \ess \inf_{x\in\R^d} V_o(x)$ we have
\[H_{\omega,L}\ge -\mu G\Delta_L G-c.\]
Hence by min-max principle
\[\tr{\id_{(-\infty,E]}\pa{H_{\omega,L}}}\le \tr{\id_{(-\infty,E+c]}\pa{-\mu G\Delta_L G}= \tr{\id_{[-\kappa,\kappa]}\pa{-\mu U\Delta_L U^*}}}\]
for $E<\infty$, where $U=U^*:=G^{-1}_- G$ and $\kappa:=\pa{E+c}G_-^{-1}$. Since $S:=U^{-1}$ satisfies $\|S\|\le1$, we are now in position to conclude via Lemma \ref{lem:SpecCong} that
\[\tr{\id_{(-\infty,E]}\pa{H_{\omega,L}}}\le \tr{\id_{[-\kappa ,\kappa]}\pa{-\mu\Delta_L }}\le C_{E,\mu} L^d,
\]
where the latter bound is well known \cite{Sim}.
\end{proof}

\begin{lem}[Wegner estimate]
\label{lem:WegnerDef}
 For every $E>0$ there exists $C_{\W}= C_{\W,E}$ such that for all $I\subset (-\infty,E]$
\beq
\label{def:Wegner'}
\P\pa{\tr \id_{I}(H_{\omega,L})\geq 1} \leq C_{\W} L^d \abs{I}.
\eeq
\end{lem}
\begin{proof}
Let $I=\mce +[-\delta,\delta]$ for suitable $\mce<E$ and $\delta>0$.  Using  $ \tr \id_{I}(H_{\omega,L})= \tr \id_{[-\delta,\delta]}(H_{\omega,L}-\mce)$ and Lemma \ref{lem:SpecCong} we get 
\beq
\tr \id_{[-\delta,\delta]}(H_{\omega,L}-\mce)\le \tr \id_{[-\delta,\delta]}(S\pa{H_{\omega,L}-\mce}S^*),
\eeq
where $S=G_-G^{-1}$. If we introduce the auxiliary periodic potential $\wt V_{o,\mce}:=G_-^2G^{-2}V_o-\mce G_-^2G^{-2}$ and the random potential $\wt V_\omega:=G_-^2G^{-2}V_\omega$, then
\[\wt H_{\omega,L}:=S\pa{H_{\omega,L}-\mce}S^*=-\mu G_-^2\Delta+\wt V_{o,\mce}+\wt V_\omega\]
is a standard ergodic RSO for which the Wegner estimate is known. The statement follows since the constant for Wegner's estimate at energy zero can be chosen to be stable in the norm of the periodic background potential. This can for instance be seen from \cite[Theorem 2.4]{CGK1}. As mentioned in the proof of Theorem \ref{cor4}, the proof from \cite{CGK1} extends to Dirichlet boundary conditions.
\end{proof}

\section{Eigenfunction decay for localized energies}
For standard RSO $H_\omega:= -\mu\Delta+V_\omega$ as in Section \ref{sec1} we briefly sketch the proof of Lemma \ref{cont:lem5}. The exponential decay of eigenfunctions in the localized regime that it describes is a direct consequence of the bound \eqref{def:LocFMB} and the Wegner estimate. 

As before, we denote $\Lambda_\ell^L(x) := \Lambda_\ell(x) \cap \Lambda_L$ for $L\geq \ell$ and $x\in\Lambda_L$.  For a set $S\subset\R^d$, we will use the notation $\partial S$ for its topological boundary. For $U\subset\Lambda$ we set $\partial^L_1 U :=\set{u\in U:\, \dist\pa{u,\partial U\setminus\partial\Lambda_L}\le1}$.

\begin{lem}
\label{SomeAuxLem2}
Let $J\subset\R$ an interval and assume that $H_{\omega}$ satisfies \eqref{def:LocFMB} for all $E\in J$. Then there exist $\tilde m,\mell_{\loc}>0$ such that for $L\geq \ell \geq \mell_{\loc}$, with probability $\geq 1- L^{2d}e^{-\tilde m\ell}$ the following holds: For all $\lambda$ in $J$ and all $x,y\in\Lambda_L$ that satisfy $|x-y|\geq \ell+2R$
\begin{align}
&\text{ either } \big\| \chi_{y}\big( H_{\omega,\Lambda^L_{\ell}(y)}-\lambda\big)^{-1}\chi_{\partial^L_1\Lambda_\ell^L(y) } \big\|\le e^{-\tilde m\ell}\label{AuxLemeq1}\\
&\text{ or } \big\| \chi_{x}\big( H_{\omega,\Lambda^L_{\ell}(x)}-\lambda\big)^{-1}\chi_{\partial^L_1\Lambda_\ell^L(x)} \big\|\le e^{-\tilde m\ell}\label{AuxLemeq2}.
\end{align}
\end{lem}
\begin{proof}
For  the lattice case this assertion has been proven in \cite[Proposition 5.1]{ETV}. The proof immediately extends to the continuum case, as, in addition to \eqref{def:LocFMB}, it only relies on the  Wegner estimate, Lemma \ref{lem:WegnerDef}. 
\end{proof}
\begin{lem}\label{SomeAuxLem}
Let $\omega$ be a configuration for which the conclusion of Lemma \ref{SomeAuxLem2} holds. Then  for all $\lambda\in J$ there exists $x=x_\lambda\in \Lambda_L$ such that for all $y\in \Lambda_L\setminus\Lambda^L_{2\ell+4R}(x)$ we have
\begin{equation}
\label{stat:SomeAuxLem}
\big\| \chi_{y}\big( H_{\omega,\Lambda^L_{\ell}(y)}-\lambda\big)^{-1}\chi_{\partial^L_1\Lambda_\ell^L(y) } \big\|\le e^{-\tilde m\ell }.
\end{equation}
\end{lem}
\begin{proof}[Proof of Lemma \ref{SomeAuxLem}]
We have two possibilities: Either we can find some $x\in\Lambda_L$ such that \eqref{AuxLemeq2}  does not hold, or there is no such $x$. In the first one the assertion (with the same choice of $x$) immediately follows from \eqref{AuxLemeq1}; in the second case we can choose $x$ arbitrary.
\end{proof}

The next assertion  is used in  the proof of Lemma \ref{thm:MainResult}.
\begin{lem}
\label{cont:lem5}
Let $\omega$ be a configuration for which the conclusion of Lemma \ref{SomeAuxLem2} holds. Then, given an eigenpair $(\lambda,\psi)$ of $H_{\omega,L}$ with $\lambda\in J$, there exists $x=x_\lambda\in\Lambda_L$ such that with $m':=\tilde m/2$
\begin{enumerate}
\item[(i)] $\|\psi\|_y \leq  e^{- m'\ell}$ for all $y\in\Lambda_{L}$ with $|x-y| \geq \ell+2R$,
\item[(ii)]$\dist\big(\sigma(H_{\Lambda^L_{2\ell+4R}(x)}),\lambda\big)\leq e^{- m'\ell}$.
\end{enumerate}
\end{lem}

\begin{proof} 
Part (i):
 Let $x$ be as in Lemma \ref{SomeAuxLem} and let $y\in\Lambda_L\setminus\Lambda^L_{2\ell+4R}(x)$. By $\sigma_\ell$ we will denote a smooth characteristic function of $\Lambda^L_\ell(y)$, i.e. a smooth function with $\chi_{\Lambda^L_{\ell-1}(y)} \leq \sigma_\ell \leq \chi_{\Lambda^L_{\ell}(y)}$ and $\|\partial_i\sigma_\ell\|_\infty, \|\partial_{i,j}\sigma_\ell\|_\infty \leq 4$ for $i,j\in\{1,\ldots,d\}$. Since 
\beq\label{eq:GRI}
[H_{\omega,L},\sigma_\ell]=H_{\omega,\Lambda^L_\ell(y)}\sigma_\ell-\sigma_\ell H_{\omega,L},
\eeq
we obtain the identity
\beq
\chi_{y}\big( H_{\omega,\Lambda^L_\ell(y)}-\lambda\big)^{-1}[ H_{\omega,L},\sigma_\ell]\psi =\chi_{y}\psi.
\eeq
Together with $[ H_{\omega,L},\sigma_\ell]=\chi_{\partial^L_1\Lambda_\ell(y) }[ H_{\omega,L},\sigma_\ell]$ this implies
\beq\label{eq:psi_y}
\|\psi\|_y=\norm{ \chi_{y}\psi}\le \Big\| \chi_{y}\big( H_{\omega,\Lambda^L_\ell(y)}-\lambda\big)^{-1}\chi_{\partial^L_1\Lambda^L_\ell(y) }\Big\|\cdot
\norm{[ H_{\omega,L},\sigma_\ell]\psi}.
\eeq 
To bound the first factor on the right hand side, we use \eqref{stat:SomeAuxLem}.
For the second term in \eqref{eq:psi_y} we express
\beq
[H_{\omega,L},\sigma_\ell]\psi=-[\Delta_L,\sigma_\ell]\psi=-\pa{\lambda-\lambda_0}[ \Delta_L,\sigma_\ell]\pa{ H_{\omega,L}-\lambda_0}^{-1}\psi
\eeq
with  
\beq\label{eq:lambda0}
\lambda_0=\inf\sigma(H_{\omega,L})-1.
\eeq The statement  now follows from the bound 
\begin{align}
\label{pf:cont:lem5eq1}
\norm{[\Delta_L,\sigma_{\ell}]\pa{ H_{\omega,L}-\lambda_0}^{-1} }\le C,
\end{align}
see, e.g., \cite{Sim}.

Part (ii): For the proof we abbreviate $\tilde\ell := 2\ell+4R$. By $\sigma_{\tilde\ell}$ we will denote a smooth characteristic function of $\Lambda^L_{\tilde \ell}$. Applying the analogue of \eqref{eq:GRI} on the eigenfunction $\psi$, we get 
\beq\label{eq:commapr}
[H_{\omega,L},\sigma_{\tilde\ell}]\psi=\big(H_{\omega,\Lambda^L_{\tilde\ell}(x)}-\lambda\big)\sigma_{\tilde\ell}\psi.\eeq
We claim that the left hand side is bounded in norm by $e^{-\tilde m\ell/2}$. This implies that the function $\sigma_{\tilde\ell}\psi$ is an approximate solution of $\big(H_{\omega,\Lambda^L_{\tilde\ell}(x)}-\lambda\big)f=0$. Combining this observation with the bound $1\ge \|\sigma_{\tilde\ell}\psi\|\ge 1-L^de^{-\tilde m\ell}$ that follows from Part (i), 
we deduce Part (ii) (cf. \cite[Lemma 3.4]{EK} and its proof). 

Let $\wt\sigma_{\tilde\ell}$ be a smooth function such that $\chi_{\supp\nabla\sigma_{\tilde\ell} }\leq \wt\sigma_{\tilde\ell} \leq \chi_{\Lambda^L_{\tilde\ell+1}(y)\setminus \Lambda^L_{\tilde\ell-2}(y)}$ and such that $\|\partial_i\wt\sigma_{\ell}\|_\infty, \|\partial_{i,j}\wt\sigma_{\ell}\|_\infty \leq 4$ for $i,j\in\{1,...,d\}$.

To establish the claim, we first express (a multiple of) the left hand side of \eqref{eq:commapr} as
\begin{align}\pa{\lambda-\lambda_0}^{-1}[H_{\omega,L},\sigma_{\tilde\ell}]\psi
&=[H_{\omega,L},\sigma_{\tilde\ell}]\wt\sigma_{\tilde\ell}\big(H_{\omega,L}-\lambda_0\big)^{-1}\psi \notag\\  &= [H_{\omega,L},\sigma_{\tilde\ell}]\big(H_{\omega,L}-\lambda_0\big)^{-1}\wt\sigma_{\tilde\ell}\psi \notag\\ &\,+\pa{\lambda-\lambda_0}^{-1}[ H_{\omega,L},\sigma_{\tilde\ell}]\big(H_{\omega,L}-\lambda_0 \big)^{-1}[ H_{\omega,L},\wt \sigma_{\tilde\ell}]\psi,\end{align}
with $\lambda_0$ is given in \eqref{eq:lambda0}. 
We can bound the first term on the right hand side by 
\[\Big\|[ H_{\omega,L},\sigma_{\tilde\ell}]\big( H_{\omega,L}-\lambda_0 \big)^{-1}\Big\|\big\|\chi_{\Lambda^L_{\tilde\ell+1}(y)\setminus \Lambda^L_{\tilde\ell-2}(y)}\psi\big\|\le C\pa{\tilde\ell+1}^de^{-\tilde m\ell}\le \frac{e^{-\tilde m\ell/2}}{2}\]
by Part (i). The second term can be bounded by  
\[\pa{\lambda-\lambda_0}^{-1}\Big\|[H_{\omega,L},\sigma_{\tilde\ell}]\big(H_{\omega,L}-\lambda_0 \big)^{-1} [H_{\omega,L},\wt \sigma_{\tilde\ell}] \Big\|\big\|\chi_{\Lambda^L_{\tilde\ell+1}(y)\setminus \Lambda^L_{\tilde\ell-2}(y)}\psi\big\|\le \frac{e^{-\tilde m\ell/2}}{2}\]
as well, and the result follows.
\end{proof}

\end{appendix}

\bigskip
\footnotesize
\noindent\textit{Acknowledgments.}
A.E. was partly supported by the  Simons Foundation (grant \#522404). A.D. is very grateful to Jean-Claude Cuenin, Peter M\"uller, and Ruth Schulte for illuminating discussions.

\end{document}